\documentclass[final]{svjour2}
\usepackage{graphicx}
\usepackage{rotating}
\usepackage{amssymb}
\usepackage{mathptmx}
\usepackage[numbers]{natbib}
\usepackage{dcolumn}
\usepackage{bm}
\usepackage[normalem]{ulem}
\makeatletter
\journalname{}

\makeatletter 
\DeclareRobustCommand\onlinecite{\@onlinecite} 
\def\@onlinecite#1{\begingroup\let\@cite\NAT@citenum\citealp{#1}\endgroup} 
\makeatother

\begin{document}

\title{Quantum Fluids in Nanotubes: a Quantum Monte Carlo
Approach}

\author{M. C. Gordillo$^1$ \and J. Boronat$^2$ }

\institute{$^1$ Departamento de Sistemas F\'{\i}sicos, Qu\'{\i}micos
y Naturales, Facultad de Ciencias Experimentales, Universidad Pablo de
Olavide,
Carretera de Utrera, km 1, 41013 Sevilla, Spain. 
\email{cgorbar@upo.es}\\
$^2$ Departament de F\'{\i}sica i Enginyeria Nuclear,
Universitat Polit\`ecnica de Catalunya,
B4-B5 Campus Nord, 08034 Barcelona, Spain.
\email{jordi.boronat@upc.edu}
}

\date{\today}

\maketitle

\begin{abstract}

We review quantum Monte Carlo results on energetic and structure properties
of quantum fluids adsorbed in a bundle of carbon nanotubes. Using realistic
interatomic interactions the different adsorption sites that a bundle offer
are accurately studied and compared in some cases with strictly
one-dimensional geometries. The study is performed quite extensively for
$^4$He and restricted to the inner part of a single nanotube for H$_2$ and
D$_2$. From a theoretical point of view, nanotubes open the real
possibility of a quasi-one-dimensional confinement where to study quantum
fluids in extremely reduced dimensionality. The results obtained show that
in the narrowest configurations the system is nearly one-dimensional
reinforcing the interest on the physics of one-dimensional quantum fluids.
Experimental confirmation of the theoretical results obtained is still not
in a satisfactory situation due to the difficulties on extracting from the
data the dominant adsorption sites.

\keywords{ Nanotubes, Adsorption, Quantum Monte Carlo}

\PACS{ 74.70.Tx,74.25.Ha,75.20.Hr}

\end{abstract}

\section{Introduction}
Carbon nanotubes were discovered by Iijima \cite{Iijima} in 1991. 
They are the result of taking one or several graphene sheets and 
roll them up to form hollow seamless tubes whose common characteristic is to 
have a very large aspect ratio \cite{harris,reich}. When a single graphene
layer is used, we have what it is called Single Walled Carbon Nanotubes
(SWCN) in opposition to the Multiple Walled Carbon Nanotubes (MWCN)
that are structures formed by two or more nanotubes nested with each other.     
In this work, we will be concerned with the absorption properties of 
SWCN, since the corresponding to MWCN tubes are expected to be similar. 
The typical form of obtaining a single walled tube is by laser ablation, 
high pressure CO conversion or with an 
arc-discharge technique. With those, one can obtain a
gaussian distribution in the diameters of the tubes, the average being
between 1 and 1.5 nm \cite{pro1,pro2,pro3}. Single walled tubes tend
to associate to each other to form bundles of triangular lattice-like
sections. 

The previous prescription to create a nanotube indicates that its main
properties are the diameter and what it is called the chiral angle. 
This angle is related to the relative orientation of the graphene sheet, that
is used to form the nanotube, and the main axis of the tube formed. 
Given the close relations between a graphene sheet and a carbon nanotube, 
those are defined in terms of the graphene vectors 
${\bf a}_1$ and  ${\bf a}_2$. 
These are the shortest ones connecting the center of any given
hexagon to the center of two other neighboring hexagons.  
This definition implies that 
$|{\bf a}_1| = |{\bf a}_2|$ = 2.46 \AA \ with an angle between them 
of 60$^0$. In the unit cell so defined, we have two carbon atoms at
the positions $\frac{1}{3} ({\bf a}_1 + {\bf a}_2)$ and
$\frac{2}{3} ({\bf a}_1 + {\bf a}_2)$. To form a nanotube, we need
what it is called a chiral vector ${\bf c}$ of the type 
${\bf c}$ = $n {\bf a}_1  + m {\bf a}_2$ and identify its length with 
the circumference of the desired tube, that in turn will be named 
($n,m$).

A bundle of carbon nanotubes offers a variety of adsorption places that
hardly can be found in other structures. Apart form the inner part of a
single nanotube, that can have different diameters, more room for
adsorption appears in the narrow interchannels between three neighboring
tubes and in the grooves that appear in the external surface of the bundle
\cite{coleRMP}. It is particularly interesting from a fundamental point of
view the possibility of playing with a nearly one-dimensional confinement.
If, in addition, one is interested in quantum fluids then
nanotubes afford the unique and fascinating possibility of the theoretical
study of nearly one-dimensional quantum fluids. In some sense, this is
similar to the new physics that emerged thirty years ago 
from the experimental observation
of films of quantum fluids (quasi two dimensions) adsorbed on graphite 
and other planar substrates. In this article, we review recent theoretical
results on the physical adsorption of $^4$He, H$_2$, and D$_2$ on the
different sites that a bundle makes possible. 
The study has been carried out using realistic
interactions and the diffusion Monte Carlo method that allows for a very
accurate description of the ground state of the system. Unfortunately, a
great deal of our theoretical results cannot be trusted with available
experimental data due to the difficulties of a reliable identification of
the dominant adsorption places. However, there is some agreement on the
fact that the gases are predominantly adsorbed on the external surface of
the bundle. As we report in Section 3, the presence of defects in the
carbon nanotube may partially explain the low adsorption rate of helium or
hydrogen in its inner surface.      

The rest of the work is organized as follows. In Sec. 2, we present the
diffusion Monte Carlo method used in our microscopic study and describe the
interactions present in the Hamiltonian and the trial wave functions used
for importance sampling. Results of $^4$He adsorbed in different sites of a
nanotube bundle are presented in Sec. 4 and the ones for H$_2$ and D$_2$
are comprised in Sec. 5. Finally, some concluding remarks are contained in
Sec. 6.

\section{Method}
Our goal in this study was to obtain the ground state of different
quantum liquids when they are confined inside a nanotube or  close to 
the  external surface  of both a single nanotube and a bundle of them. 
To do so, we employed a quadratic diffusion Monte Carlo method (DMC) \cite{boro94}. 
Nowadays, this is a 
standard technique to solve the Schr\"odinger equation of a many-body 
system using a stochastic approach. The starting point is the equation 
written in imaginary time $t$ (in units of $\hbar$),
\begin{equation}
-\frac{\partial \Psi({\bf R},t)}{\partial t}  = (H - E)  \Psi({\bf R},t) 
\end{equation}
where $ \Psi({\bf R},t)$ is the wave function of the $N$-particle system, 
and ${\bf R}$ stands for a set of $3 N$ coordinates for all the $N$  
atoms considered in the simulation.  Written in that way, and 
taking into account that 
\begin{equation}     
H = -\frac{\hbar^2}{2m}  \sum_{i=1}^{N} \nabla_i^2 + V({\bf R})
\end{equation}
the above Eq. 1 can be considered as a diffusion equation and treated as
such. However, the variance of the results is greatly reduced when an
importance sampling scheme is introduced. To do so, one solves the
Schr\"odinger equation for the wave function 
\begin{equation} 
f({\bf R},t) = \Psi({\bf R},t) \Phi({\bf R})    \ ,
\end{equation}
where $\Phi({\bf R})$ is a time-independent trial function. In this
function one introduces relevant information known {\em a priori} about the system. 
For instance, if
we expect the system to be confined in a given portion of space, 
$\Phi({\bf R})$ should have a maximum around that region and be close to zero 
elsewhere. The Schr\"odinger equation for the wave function $f({\bf R},t)$
turns out to be   
\begin{eqnarray}
-\frac{\partial f({\bf R},t)}{\partial t} &  =  &  -D {\mbox{\boldmath $\nabla$}}^2 
f({\bf R},t) + D {\mbox{\boldmath $\nabla$}} ({\bf F} \, f({\bf R},t)) 
 + (E_{\rm L}({\bf R}) -E) f({\bf R},t)  \nonumber \\ 
 & \equiv & (A_1 + A_2 + A_3)  f({\bf R},t)  \label{dmc1} \ ,
\end{eqnarray}
where $D = \hbar^2 / 2m$, and $A_1$,  $A_2$ and  $A_3$ are the three
operators acting on $f({\bf R},t)$ in the sum (first line of Eq.
\ref{dmc1}) in the same order as they appear. The local energy,  
\begin{equation}
E_{\rm L}({\bf R}) = \Phi({\bf R})^{-1} H \Phi({\bf R}) 
\end {equation} 
serves as an estimator for the energy of the considered system, and  the term 
\begin{equation}
{\bf F} ({\bf R}) = 2 \Phi({\bf R})^{-1} \nabla  \Phi({\bf R})
\end{equation} 
is the drift force introduced by the importance sampling. 
In practice, this term
allows for sampling  mainly the regions in which the trial function
has greater values and thus the efficiency of the method is increased.    

To solve Eq. 4 it is transformed to integral form,
\begin{equation}
f({\bf R}',t+\Delta t)) = \int G({\bf R}',{\bf R},\Delta t) f({\bf R},t)
d{\bf R} \ ,
\end{equation}
with $G({\bf R}',{\bf R},\Delta t)= \langle {\bf R}' | \exp(-\Delta t \, H) |
{\bf R} \rangle$ the Green function. When $\Delta t \rightarrow 0$, it can
be well approximated by the short-time approximation
\begin{eqnarray}
\lefteqn{f({\bf R}',t+\Delta t)) =} \\
& & \int \exp [- \frac{\Delta t}{2} A_3] \exp [- \frac{\Delta t}{2} A_2]
 \exp [- \Delta t A_1]  \exp [- \frac{\Delta t}{2} A_2]  
\exp [- \frac{\Delta t}{2} A_3]  f({\bf R},t) d{\bf R} \nonumber
\end{eqnarray}
The partition described in the equation 
above is not unique, but this form assures that the errors
due to the short-time approximation for the Green function are 
of the order $(\Delta t)^2$.

In the diffusion Monte Carlo method, $f({\bf R},t)$ is not represented
by an analytical function, but by $M$ sets of $N$ particle coordinates,
${\bf R}$. Any 
of those sets is called a {\em walker}. This means that to apply the 
equation above to $f({\bf R},t)$, we have to modify any of the $N$
coordinates of the $M$ walkers by means of the following prescription:        

(a) Change the coordinates of the particles in each walker by means of the 
drift force an interval $\Delta  t/2$, i.e.,
\begin{equation}
{\bf R}' = {\bf R} + D {\bf F} ({\bf R}) (\Delta t)/2
\end{equation} 

(b) Change again the coordinates by adding to each of them a random number 
$\chi$ drawn from a gaussian distribution in three dimensions 
$\exp [-\chi^2/(4 D \Delta t)]$
\begin{equation}
{\bf R}'' = {\bf R}' + \chi
\end{equation}

(c) Repeat step (a)

(d) Calculate the factor
\begin{equation}
n_s = \exp [ - \Delta t (E_{\rm L}({\bf R}) -E)]
\end{equation} 
for each walker ${\bf R}$ and replicate it $n_s$ times. This 
produces a new set of $M'$ walkers in which to perform the whole process 
again. The procedure is repeated as many times as needed to reach convergence 
to the limit $t \rightarrow  \infty$. Once that limit is 
reached, the expectation values of any of the observables could be obtained
from the set of walkers derived from the process outlined above. In 
particular, as already mentioned, the estimator for the energy is 
the local energy (Eq. (5)). The quality of the results 
could be judged by the variance of the local energy, that depends directly
on the trial wave function. If $\Phi({\bf R})$ is the exact solution of the
Schr\"odinger equation for the system under consideration, the variance 
of the energy is exactly zero. If not, the DMC technique is able to obtain
the exact ground state for a system of bosons within some statistical
noise.

To apply the diffusion Monte Carlo method to a particular system, we need then a
set of walkers ${\bf R}$, a trial function, and an interaction potential
between the particles. As indicated above, the trial
function  incorporates information known {\em a priori} about the 
system. In particular, $\Phi({\bf R})$ avoids configurations with two particles
(in our case two atoms or molecules in the quantum liquid) 
sharing the same position. This is traditionally made with a Bijl-Jastrow
wave function,      
\begin{equation}
\Phi_J = \prod_{i<j} f(r_{ij})
\label{jastrow}
\end{equation}
where $r_{ij}$ is the distance between two $^4$He atoms or two H$_2$ or
D$_2$ molecules, depending on the case. The two-body correlation function 
$f(r)$ goes to zero when the
distance between particles is very small and approaches one in bulk systems
at large interparticle separations. A simple and very efficient model for
$f(r)$ is provided by the McMillan function
\begin{equation}
f(r) = \exp \left[-\frac{1}{2} \left(\frac{b}{r} \right)^5 \right] \ ,
\label{millan}
\end{equation}  
with a parameter $b$ that can be 
optimized in a separate variational Monte Carlo (VMC) calculation, 
and that is different for each species adsorbed in the 
nanotubes.

When a quantum liquid is adsorbed on a surface one has also to include
in the Hamiltonian interactions between the atoms and the substrate. 
Also, the trial wave function for importance sampling has to 
 avoid situations in which the 
adsorbed species is too close to the nanotube. This is usually made by 
multiplying the Jastrow factor $\Phi_J$ (\ref{jastrow}) by another function, 
$\Phi_C$. For that, we have 
several possibilities. One of them is to consider the nanotube as a 
confining structure that interacts with a given particle as a whole.   
In that case, we can use  
\begin{equation}
\Phi_C = \prod_{i} \Phi_i  \ ,
\end{equation}
$\Phi_i$ being the result of solving the Schr\"odinger equation for 
a single particle ($^4$He,H$_2$ or D$_2$) under the potential defined
by the entire nanotube. A second option consists in using an educated guess. 
For instance, if the tube is narrow enough (as in a (5,5) case), $\Phi_i$     
could be taken  as a Gaussian
\begin{equation}
\Phi_i = \exp [-a (x_i^2 + y_i^2)]
\end{equation}
where $a$ is a variational parameter to be properly optimized, and  
$x_i$ and $y_i$ are the coordinates of the $i$-atom or molecule with respect to
the center of the tube. The use of this expression in a (5,5) case, with a
proper $a$, provides a trial function of high quality, as can be 
seen in the corresponding energy variance (see below).  

The third possibility is to use a Jastrow model similar to $\Phi_J$
(\ref{jastrow}) but with a value for
$b$ adapted to the C-He interaction. In this case,   
\begin{equation}
\Phi_i = \prod_j \exp \left[-\frac{1}{2} \left(\frac{b}{r_{j-i}} \right)^5 \right]
\end{equation}
where $j$ indicates here the positions of all the carbons in the corresponding
nanotube or bundle.  In any case, the whole trial wave function is the  product
$\Phi_J \Phi_C$. Occasionally, $\Phi_C$ could be the product of the forms
considered in Eqs. 15 and 16.  

All the above indications are for a liquid system, i.e., for a system in which
the adsorbed species do not tend to be located around fixed positions. 
However, if we consider a solid, a realistic description should define  
the trial wave function as the product $\Phi_J \Phi_C \Phi_S$ where    
\begin{equation}
\Phi_S = \prod_{i} \exp \left[-a_s (x_i-x_{\rm site})^2 -a_s(y_i-y_{\rm site})^2 -
b_s (z_i-z_{\rm site})^2 \right] \ .
\end{equation}
The coordinates $x_{\rm site}$, $y_{\rm site}$, and $z_{\rm site}$ are the lattice 
positions in the crystal structure, and are different for each atom or molecule in the
quantum solid. The parameters $a_s$ and $b_s$ can be optimized in the same
way than the ones in previous $\Phi$'s. This form of the trial
function (Nosanow-Jastrow) can also be used when we want to localize the adsorbed species 
in some way, not necessarily according to a crystal pattern.   

The last required microscopic input to describe the systems we are dealing with is 
the interaction potential between the different species inside or around
the nanotube. This whole interaction is the $V({\bf R})$ term in the
Schr\"odinger equation (2), and consists of two parts. The first one is 
the interaction between any pair of adsorbate atoms or molecules.
In all cases described in the present work, the He-He potential was taken 
from Ref. \onlinecite{aziz}, while the  
H$_2$-H$_2$ interaction was described by the Silvera potential \cite{silvera}. 
The second part is the tube (or set of tubes)-adsorbate potential. 
To define it, we can consider the nanotube as a whole, or simply sum up
all the individual carbon-adsorbate contributions. The former case is 
simply the result of averaging over the latter, and has the advantage 
of being computationally less expensive than to take into account all
the carbon atoms in their particular positions. However, it has the 
important drawback of considering as equal tubes that have the same radius 
but different $(n,m)$ indexes. The most common choice
 for the individual C-He \cite{cole} and 
C-H$_2$ interactions \cite{cole2} is the  Lennard-Jones (LJ) model. 
The averaged potential using
LJ for a single tube-adsorbate interaction has the form \cite{uptake}
\begin{equation}
V(r,R) = 3  \pi \theta \epsilon \sigma^2 \left[ \frac{21}{32} \left(
\frac{\sigma}{R} \right)^{10} M_{11} (x) - 
\left(\frac{\sigma}{R} \right)^4 M_5(x) \right]   \ , 
\end{equation}     
where $\theta$ is the surface carbon density in a graphene sheet 
(0.38 \AA$^{-2}$), $R$ is the radius of the cylinder, and  $x = r/R$ with 
$r$ the distance to the center of the tube. The functions $M_n(x)$ are
of the form
\begin{equation}
M_n(x) = \int_0^\pi d \phi \frac{1} {( 1 + x^2 - 2 x \cos \phi)^{n/2}} \ . 
\end{equation}

In the remaining of this review, we  
report results obtained by applying the diffusion Monte Carlo method to the problem of 
adsorption of $^4$He, H$_2$ and D$_2$ inside or outside carbon nanotubes of different
radii.  
   
\section{$^{\bf 4}$He adsorbed in carbon nanotubes} 
Bulk helium is the paradigm of a quantum liquid, basically because 
it remains liquid  at zero pressure all the way to 0 K. This is the reason
why the studies of how a quantum liquid behaves when confined inside or
close to a nanotube have been made primarily with $^4$He. In this section, 
we will describe many-body DMC calculations performed in those
environments.

\begin{figure}[b]
\centering {\includegraphics[width=0.7\linewidth]{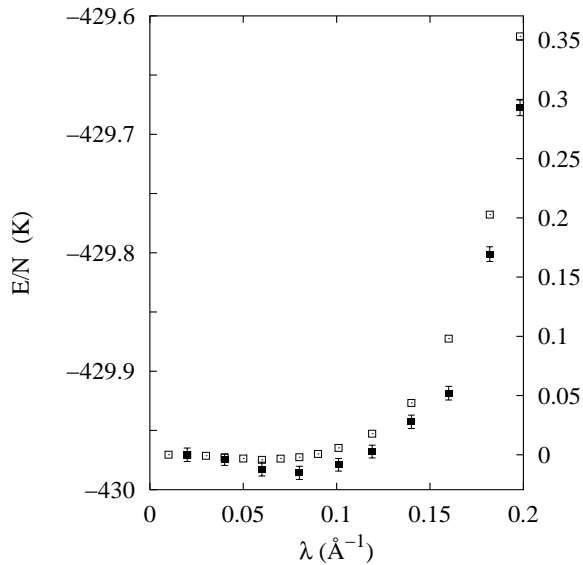}}
\caption{
Energy per particle ($E/N$) versus the linear concentration ($\lambda$),
for a strictly one dimensional system
(open squares, right energy scale), and a (5,5) nanotube (full squares,
left energy scale). 
}
\label{fig1}
\end{figure}

The most extreme form of confinement that one can impose upon a set of
particles is to force them into a one-dimensional (1D) array. 
Since, as mentioned above, a nanotube is essentially a long cylinder, this
is easily obtained by choosing a tube with small enough radius.
Obviously, the nanotube should be opened in some way, say mechanically
or chemically.   
Since curving the graphene sheet to create a nanotube imposes 
a certain degree of tension upon the C-C bonds the narrowest
stable isolate nanotube of the (n,n) type is the (5,5) one, with a radius 
of 3.42 \AA. This is the distance between the center of the carbon atoms
in the rolled up graphene layer and the center of the structure. 
Since the Lennard-Jones $\sigma$ parameters for the He-He and C-He interaction
are 2.556 and 2.74 \AA, respectively, this means that inside a (5,5) tube
there is not room enough to have two He atoms sharing the same plane 
perpendicular
to the tube axis, i.e., $^4$He should behave in this environment as a 
(quasi) 1D system.     

In Fig. 1, we report DMC results for the equation of state of  
$^4$He in a purely  
1D environment and inside a (5,5) tube at low densities. For the pure 1D array
of atoms, the trial function was of Jastrow type. 
When the nanotube was included, it  
was modeled as a smooth cylinder with an averaged over C-He potential 
and a  trial function that was the result of multiplying 
$\Phi_J$ for a set of as many Gaussians as helium atoms in the simulation 
cell (see previous Section) \cite{he4}.

In that figure, we observe two $y$ scales, corresponding to the energies
per atom ($E/N$) for  $^4$He in a pure
1D system (right) and inside a smooth (5,5) tube (left). The 
abscissa is the number of atoms per unit length ($\lambda$). The 
curves have been drawn to make the lowest density point inside the tube
to coincide with the 1D value for the same linear density. This is 
basically similar to subtract from all the tube energies the binding
energy for a single  $^4$He in a tube ($429.97$ K). From Fig. 1 we 
can infer two main results. First, 
helium inside this narrow tube is not a pure 1D system since the 
equation of state is different in both cases. Second,  
the many-body ground state is a liquid in both situations since both curves present
an energy minimum corresponding to a density greater than zero. 
A third-order polynomial fit to DMC energies, 
\begin{equation}
e = e_0 +  A \left( \frac{ \lambda - \lambda_0 }{\lambda_0} \right)^2 +
B \left( \frac{ \lambda - \lambda_0 }{\lambda_0} \right)^3 \ ,
\end{equation}
allows us to obtain the parameters $\lambda_0$ (the equilibrium density)  
and $e_0$ (the energy per particle at equilibrium). These parameters are 
given in Table I. The minimum in the equation of state is so shallow that
its location and value is much dependent on the He-He interatomic
potential: using the HFHDE2 Aziz potential \cite{azizold} the equlibirium point of 1D
$^4$He is  $\lambda_0= 0.036$ \AA$^{-1}$ and $e_0=-0.0017$ K \cite{moroni}.    

\begin{table}[t]
\begin{center}
\begin{tabular}{lcc}
Parameter   &  1D $^4$He  & $^4$He in a tube \\ \hline
$\lambda_0$ (\AA$^{-1})$ & 0.062 $\pm$  0.001 & 0.079 $\pm$ 0.003 \\
$e_0$ (K)               & -0.0036 $\pm$ 0.0002 & -429.984 $\pm$ 0.001 \\
$A$   (K)               &  0.0156 $\pm$ 0.0009 & 0.048 $\pm$ 0.006 \\
$B$   (K)               &  0.0121 $\pm$ 0.0008 & 0.0296 $\pm$ 0.009 \\
$\chi^2/\nu$            &         2.2         &    0.24           \\
\end{tabular}
\caption{Parameters of Eq. 18 for 1D helium and helium inside a (5,5) 
smooth tube.}
\end{center}
\end{table}

The comparison between $\lambda_0$ and $e_0$ ($-0.018$ K
in the tube case, subtracting the infinite dilution limit) 
indicates that the energy minimum is deeper and located at a bigger   
linear density in a tube than in a pure 1D system.  
The reason is that even though the distance between first 
neighbors is similar in both environments, it is possible for the 
second neighbors inside the nanotube to be close to each other by creating a zig-zag 
structure. This small
effect would increase both $\lambda_0$ and $e_0$ and serve to create
a quasi-one-dimensional array of atoms instead of a pure 1D one. 
A behavior completely similar to this for $^4$He was found in the cases
of H$_2$ \cite{prl2000} and D$_2$  \cite{deu}, the only difference being
that these latter ones are closer to the corresponding
1D systems. Compared with $^4$He, 
 $\lambda_0$ and $e_0$ in H$_2$ and D$_2$ are larger and deeper, respectively, both in 
pure 1D systems and inside narrow tubes (see Sec. 4). This trend also follows for
Ne, of which a purely 1D DMC calculation is reported in Ref. \onlinecite{llorens1}.

\begin{figure}
\centering {\includegraphics[width=0.7\linewidth]{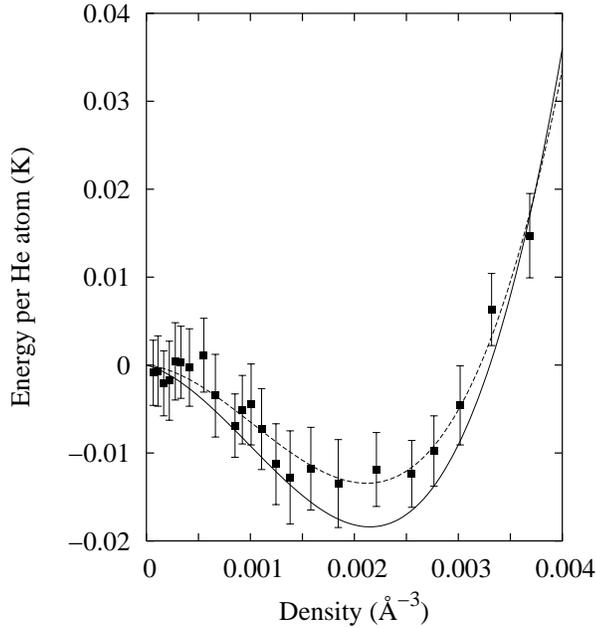}}
\caption{
Difference between the total adsorption energy and the the binding energy
for a single helium atom in a (5,5) tube. Full line, results for a smooth C-He
potential, full squares and dashed line, same for a corrugated 
C-He interaction.
}
\label{fig2}
\end{figure}

Considering a carbon nanotube as a smooth cylinder is clearly 
an approximation \cite{pro}. To check how good it is, we can make use of Fig. 2.
There, we report the equation of state for the same (5,5) tube, 
but now considering the real nanotube by taking into account all
the C-He interactions, both in the trial function and in the
potential. i.e., we consider a fully corrugated carbon nanotube. 
As in the case of Fig. 1, the infinite dilution
limits are subtracted in both curves to work in the same scale. This
is necessary since that binding energy is appreciably different in 
the smooth ($429.97$ K) and corrugated ($429.51$ K) tubes. There is also 
another difference with Fig. 1: the density in the $x$ axis is taken to 
be the volume density, i.e., $\lambda$/($\pi$ R$^2$). The results of
a fit similar to that of Eq. 20 indicates that in a corrugated structure
$\lambda_0$ is nearly the same (0.077 versus 0.079 \AA$^{-1}$ in the 
smooth tube), but $e_0$ is reduced $\sim 25$\% ($-0.013$ versus
$-0.018$ K). This means that to include the corrugation 
makes the system closer to a pure 1D one. 

Another aspect to be taken into account in this (5,5) tube is 
the possibility of having a
phase transition from a liquid to a solid phase when the density 
(and the pressure) increases, just like in the bulk case.  
Evidences of a phase transition of this kind, only possible in 1D at zero 
temperature, appear in a variational calculation of 1D $^4$He \cite{krot}. 
A solid phase is defined, as indicated above, as one in which the 
atoms are confined close to particular positions. In 
quasi-one-dimensional systems, these positions were chosen as regularly spaced 
in the $z$ direction (the one of the long axis of the nanotube).
A comparison between the DMC energies for
the liquid and solid phases is given in Table 2. One can see that in
both systems, the energy per particle when localization
is imposed ($a_s= 0, b_s \neq 0$ in Eq. 17) is below the corresponding to a 
liquid structure ($a_s = b_s =  0$) for lineal
densities greater than 0.358 \AA$^{-1}$. Unfortunately, the energy 
differences between the liquid and the solid structures at very high 
densities are not big enough to allow us to perform a Maxwell double-tangent
construction to obtain the limits of the coexistence region. This probably
means that the change is quasi continuous. In any case, this transition 
is only possible at 0 K.    

\begin{table}[b]
\begin{center}
\begin{tabular}{lcccc}
$\lambda$ (\AA$^{-1}$) &  $E/N$ (1D, liquid)  & $E/N$ (1D, solid) &
$E/N$ (T, liquid) & $E/N$ (T, solid) \\ \hline
0.406  &  123.726 $\pm$    0.012  &  123.561  $\pm$  0.012 &  -350.155 $\pm$ 0.030 & -350.20 $\pm$ 0.02 \\
0.380  &   67.070 $\pm$    0.011  &   67.000  $\pm$  0.009 &  -382.282 $\pm$ 0.016 &  -382.321 $\pm$ 0.012 \\
0.358  &   37.602 $\pm$    0.008  &   37.596  $\pm$  0.007 &  -401.873 $\pm$ 0.013 & -401.844 $\pm$ 0.010 \\
0.338  &   21.881 $\pm$    0.007  &   21.904  $\pm$  0.005 &  -413.091 $\pm$ 0.014 &  -413.061 $\pm$ 0.012 \\
0.320  &   13.240 $\pm$    0.005  &   13.258  $\pm$  0.006 &  -419.551 $\pm$ 0.011 &  -419.493 $\pm$ 0.010
\end{tabular}
\caption{Energies per particle at large $\lambda$ for the quasi one dimensional 
systems considered in the text. All the energies are  in K.}
\end{center}
\end{table}

\begin{figure}[b]
\centering {\includegraphics[width=0.7\linewidth]{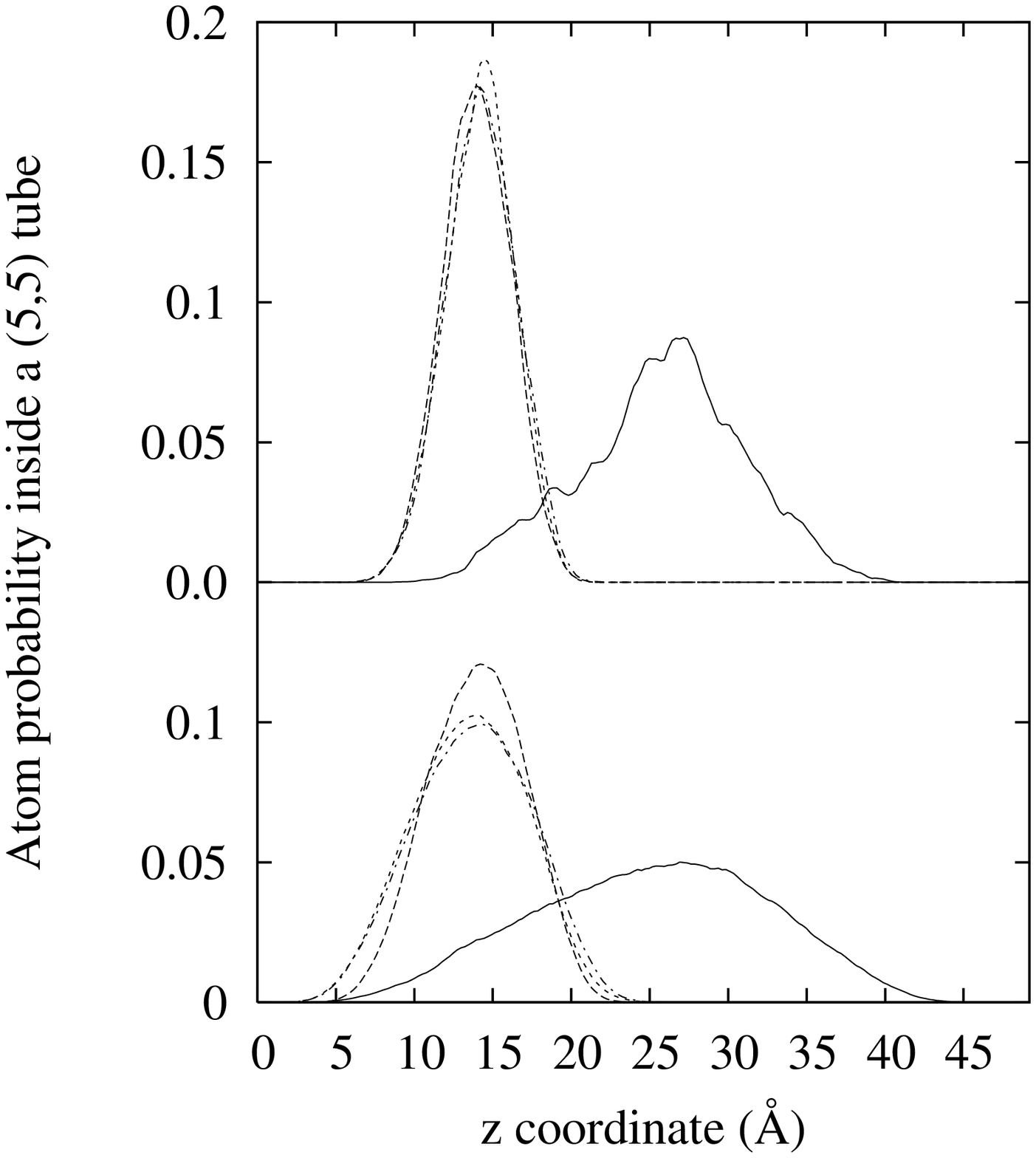}}
\caption{
Probability density of having a single atom (Ne, top; $^4$He, bottom) 
inside a given $z$ position in a (5,5) nanotube.
Full lines indicate the result for a perfect tube, dashed-dotted 
lines what we obtain when an asymmetric
5-1DB is introduced, dashed lines the situation for a symmetric defect of
the same type, and the dotted curves are the probabilities for an hypothetical
three dangling bond vacancy. 
}
\label{fig3}
\end{figure}

Apart from the corrugation, there are other influences we should take into 
account to describe the adsorption of gases in quasi 1D systems such as 
(5,5) tubes. For instance, carbon nanotubes are known to have a 
certain amount of defects \cite{teizer2,cpl,adv},  
single carbon vacancies being among the simplest of them.
In principle, the most stable
structure when a carbon atom is removed from the carbon layer conforming 
a nanotube is the so-called 5-1DB defect. This is the result of two 
of the possible dangling bonds recombining with each other to form a 
pentagon and leaving the third one unchanged.
There are two forms of doing this, termed symmetric and asymmetric.
It appears that for (n,n) nanotubes the last one is the most 
stable \cite{lu, ajayan,kra,zhang}. Diffusion Monte Carlo
calculations were carried out to check if $^4$He atoms
are allowed to enter this narrow system when a single vacancy is present
\cite{prl06}. 
To check that, a single atom of the gas was placed at $z$=0 
(the beginning of a simulation cell of length  49.2 \AA)  and left to  
evolve until no appreciable change in the
density profiles (see Fig. 3) was found. The case of $^{20}$Ne is also
contemplated. In both cases, the analysis of Fig. 3 indicates that, 
while there is no problem for those atoms to enter and explore all space
inside a defect-less (5,5), the presence of a single vacancy of any kind
would bar the loading of those tubes.

A (5,5) tube is not the only environment narrow enough to produce a 
quasi-one-dimensional system. Another possibility is the interstitial channel (IC)
located among every three carbon nanotubes when they associate to create
a bundle. A section of a bundle of three tubes is shown schematically in 
Fig. 4. This channel is even narrower than a (5,5) tube, and its 
conformation depends on the particular nanotubes that surrounds it 
\cite{yoh2}. If a perfect (with no defects) IC formed by 
three (10,10) nanotubes is considered, the same type of calculations 
made for a (5,5) tube can be performed in order to know if it is possible to 
have helium inside those IC's. However, in this case it should taken into 
account the fact that the carbon nanotubes could be displaced with respect 
to each other, creating different potential energy landscapes to be felt 
by the helium atoms. 

\begin{figure}[t]
\centering {\includegraphics[width=0.7\linewidth]{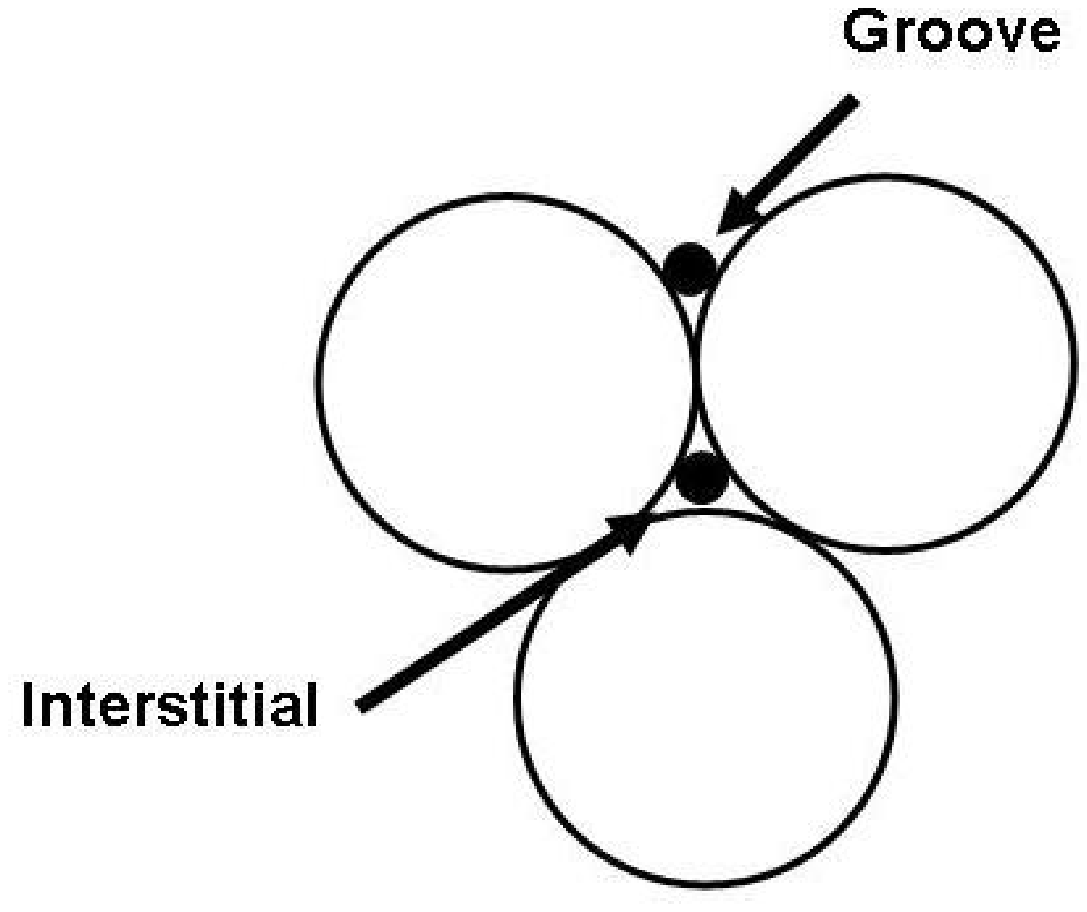}}
\caption{Sketch of a section of a bundle including three nanotubes 
together with a depiction of the location of an interstitial
and a groove adsorption locations. 
}
\label{fig4}
\end{figure}

The results are shown in Fig. 5. There, two different density profiles were 
considered for a defect-less tube. The smoother one is the corresponding
to the minimum corrugation (the configuration of the carbon nanotubes
in which the helium atoms feel the potential along the $z$ direction 
to be smoothest) and it is similar to the one of a (5,5) tube. The profile
with more structure corresponds to the opposite case of maximum corrugation.
In any case, one can observe that a $^4$He atom has no problem to enter
an IC located among three (10,10) tubes. The same behavior was found 
for H$_2$ \cite{yoh2}, for which several others possibilities were 
checked (4-tube IC's, and IC's formed by 3 different (n,n) tubes).   

\begin{figure}
\centering {\includegraphics[width=0.7\linewidth]{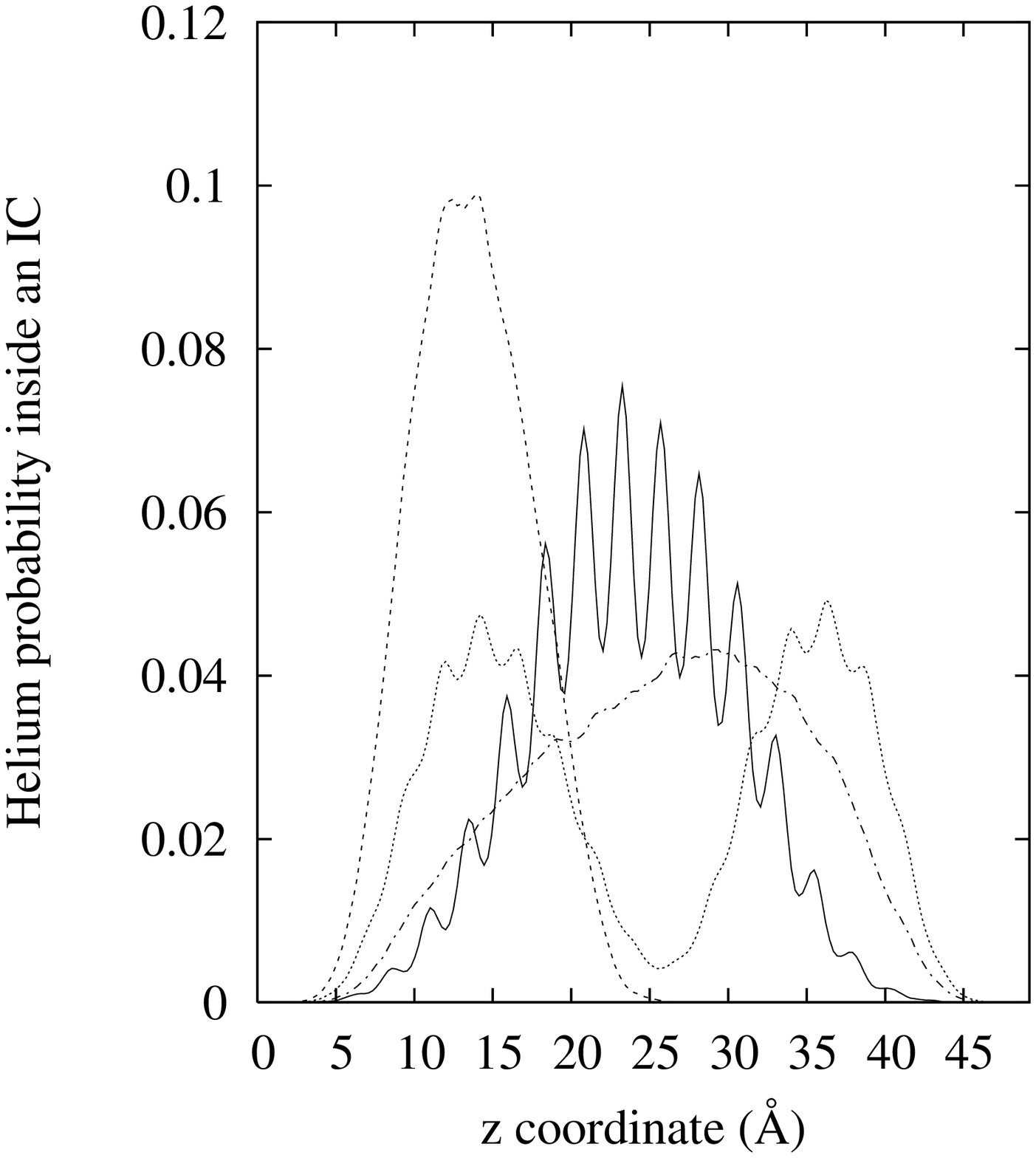}}
\caption{
Density profile for helium inside an IC. 
Solid (dashed-dotted) line, $^4$He inside a defect-free IC with the
maximum (minimum) corrugation; dashed (dotted) line, results for an asymmetric
5-1DB vacancy closest to (furthest from) the center of the IC for the
minimum corrugation cases. 
}
\label{fig5}
\end{figure}

Obviously, once helium atoms enter the defect-less IC, they can form
a quasi-one-dimensional arrangement in a similar way to the one already
considered for the  (5,5) tube \cite{bony,low}. In principle, the details of 
the equation of
state should depend on the particular kind of tubes that form the IC, 
and on their positions with respect to each other. However, in the same
way than in the narrow tube considered above, a smooth averaged tube
could be considered, in the hope that the interaction with the $^4$He atoms
is at least qualitatively similar to that of the corrugated structures. 
The trial function for a set of three smooth (10,10) tubes is also 
qualitatively different from the ones considered in Sec. 2,
\begin{equation}
\Psi({\bf R}) = \prod_{i<j}^N \exp \left[ -\frac{1}{2} \left
(\frac{b}{r_{ij}} \right)^5 \right]
\prod_i^N \prod_{n=1}^3 \exp \left[ -\frac{1}{2} \left (\frac{a}{r_{in}} \right)^5
\right] \ 
\end{equation}
i.e., it has a Jastrow-type term (the one with the double productory) that takes
care of the interaction of $^4$He  with the walls of the three tubes forming
the interchannel.

The equation of state in the low density regime is shown in Fig. 6.
There, we can see the comparison between the results of an IC, a (5,5) tube
and a pure 1D system. We can see that the energies per helium atom in the
IC are a bit 
closer to a 1D line than in the case of a cylindrical environment. A
least-squares
fit to the DMC energies in this environment with the same equation
considered previously (Eq. 20), renders  $\lambda_{0}$ = 0.076 $\pm$  0.004 
\AA$^{-1}$, and $(E - E_B)/N = 0.010$ $\pm$ 0.001 K. This means that 
the atoms inside an interchannel formed by three tubes form a 
quasi-one-dimensional arrangement that is similar to that of a (5,5) tube.    

\begin{figure}
\centering {\includegraphics[width=0.7\linewidth]{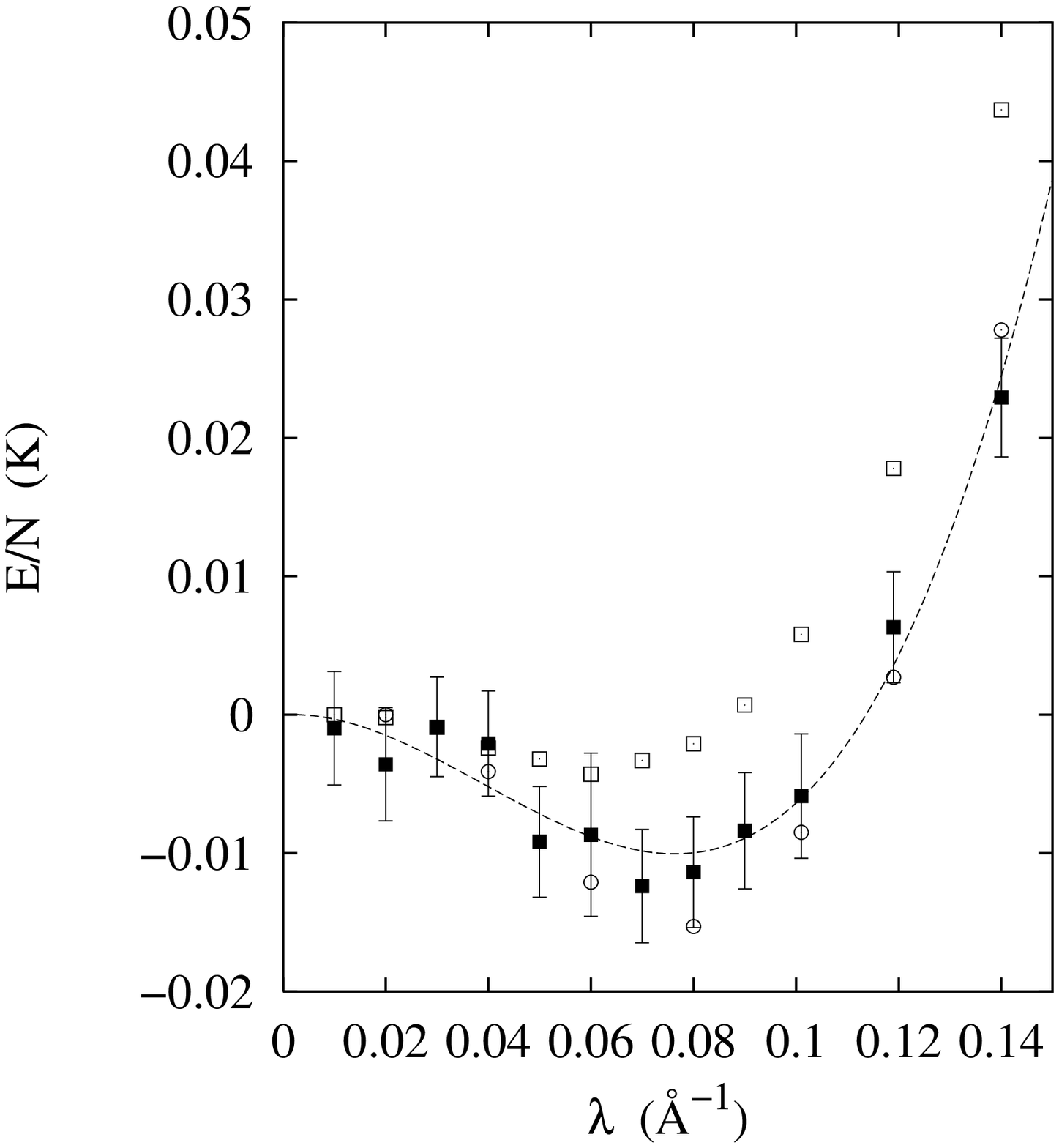}}
\caption{
Energy per $^4$He atom inside an IC formed by three smooth (10,10) tubes 
 in the density range between 0 and 0.15
\AA$^{-1}$. Open squares,  1D; filled squares,
$^4$He in a interchannel of (10,10) tubes; open circles, $^4$He in a (5,5)
tube. The dashed line is a polynomial fit 
to the IC results. The binding energy of a single $^4$He atom to the
interchannel (323.41 K) has been subtracted.
}
\label{fig6}
\end{figure}
    
In the case of interchannel adsorption there is still another issue to consider.
A bundle of carbon nanotubes is rarely formed by three tubes, what means
that it is virtually impossible to have an isolated IC in the same way
that we can have an isolated (5,5) tube \cite{vilcheslt25}. For instance, 
we can expect a certain influence in the equation of state 
of helium atoms adsorbed in neighboring 
channels of the same bundle. In a regular bundle made
of (10,10) carbon nanotubes, a simple geometrical calculation indicates 
that the minimum distance between a pair of these channels is about 
9.8 \AA, implying that the influence of filled neighboring tubes 
could be accurately described 
by a mean field approximation of the type \cite{deu,colemf}
\begin{equation}
V_{\rm mf} = \frac{\lambda}{2} \int_{- \infty}^{\infty}  d x
\ V_{\rm He-He} \left(\sqrt{x^2 + d^2}\right) \ ,
\end{equation}
with $d$ the distance between one channel and  one of its neighbors
and $\lambda$ the helium linear density.
The total energy correction is obtained by
summing up the contribution for channels with increasing $d$ values until
the desired accuracy. Thus, the total energy per $^4$He atom in the bundle
is of the form $E({\rm IC})/N + V_{\rm mf}$. The mean-field approach  
assumes that the correlation
effects between interchannels are negligible, the net influence being only
the increase of binding energy due to the van der Waals attractive tails.
This has been checked to be essentially correct \cite{deu}. The difference
between the equations of state of a single IC and a set of them is given
in Fig. 7. We can see that the overall effect is to increase the 
binding energy and the equilibrium density of the system. 
This behavior was also observed in H$_2$ 
\cite{interh2} and Ne \cite{llorens1,llorens2}.

\begin{figure}
\centering {\includegraphics[width=0.7\linewidth]{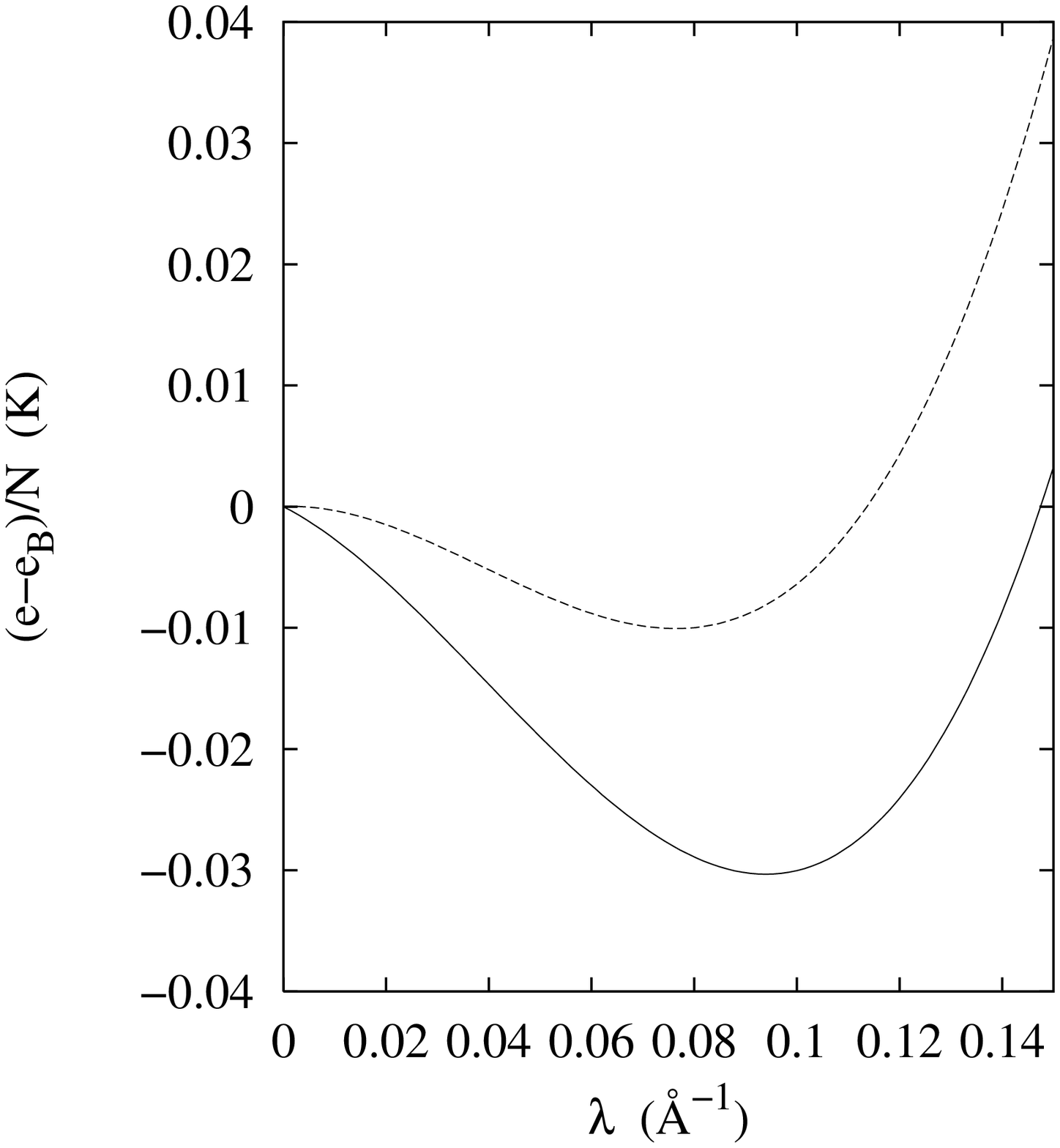}}
\caption{
Energy per $^4$He atom as a function of the linear density
$\lambda$. Dashed line, a single IC. Solid line,
a IC considering the influence of $^4$He atoms in  the surrounding
interchannels. In both cases, the binding energy of a single helium atom
to the interchannel has be subtracted. 
}
\label{fig7}
\end{figure}

In the previous calculations we have assumed that $^4$He atoms can be effectively 
adsorbed inside the interchannels and, in fact, this would be the case within 
perfect defect-less tubes. However, 
in view of Fig. 5, this could be not fully realistic. Besides the 
already discussed case of a no vacancy tube, in Fig. 5 it is displayed what happens
when a 5-1DB asymmetric vacancy (the most stable one) is introduced
in the center of a IC of a tube 49.2 \AA \ long. Two cases are considered, 
one in which the hole is directly below the trajectory of the adsorbed atom,
and another in which the defect is the one furthest away of this position. 
We can see than in the first case the helium density profile vanishes  
beyond the place in which the defect is located 
(see discussion above), what 
indicates that the adsorption in this case is prohibited. The reason is 
that once the atom reaches that position, it cannot progress further and
blocks the way to other atoms. However, when
the defect is further away, the helium atoms are able to pass the 
potential barrier, leaving a minimum in the density profile close to 
the vacancy position. Since, as mentioned above, all nanotubes are 
thought to have a certain fraction of defects, and some of them are 
just inside an IC, this means than most of the IC's will be 
at least partially empty. 

Nevertheless, if the tubes are wide enough it is sure that  
helium can enter inside them. For instance, we can see what happens
in the case to a (10,10) tube, whose radius is 6.8 \AA. Obviously, we
can only consider open-ended tubes, i.e., those whose caps have been 
removed by chemical or mechanical ways. Intuitively, one can think that 
$^4$He atoms would first go close to the wall of the 
nanotube forming a cylindrical shell. This layer creates an empty
space in the center of the tube in which additional helium atom can then enter.   
DMC calculations of this system were carried out in Ref. \onlinecite{yo3}.   
In principle, all the C-He interactions were taken into account, both 
in the potential energy and in the trial function, the latter being 
a product of the forms already described in Eq. 14 and Eq. 16. i.e.,
a product of C-He Jastrow functions and  the one-body solution of 
the Schr\"odinger equation for a smooth tube. 

Fig 8. displays results obtained for a single liquid shell (circles)
versus a single solid shell close to the wall (full squares). 
The density in the abscissae is simply the number of atoms divided 
by the volume of the tube, considering the radius as to be
the distance between the center of the tube and the center of the
graphene sheet; in this tube, 6.8 \AA. As indicated 
above, a solid is a phase in which the atoms are located 
around fixed positions. In practice, this means that the trial 
function for a solid phase includes  a set of localized Gaussians of 
the kind shown in Eq. 17. Fig. 8 suggests that at low
densities the most stable phase is a liquid, but this changes at high
enough densities. However, to determine if a single solid layer is the most
stable structure, all the possibilities have to be considered, in particular
the one where one has simultaneously a liquid layer close to the wall and an
additional liquid layer on top of it (circles in Fig. 8). From data in 
Fig 8, we can see that a single solid has very similar energy per atom
than a two-layer liquid of the same density. In addition, at higher densities
than the ones displayed in Fig. 8, another phase made of 
a solid layer closer to the wall and a liquid part in the center is observed.     

\begin{figure}
\centering {\includegraphics[width=0.7\linewidth]{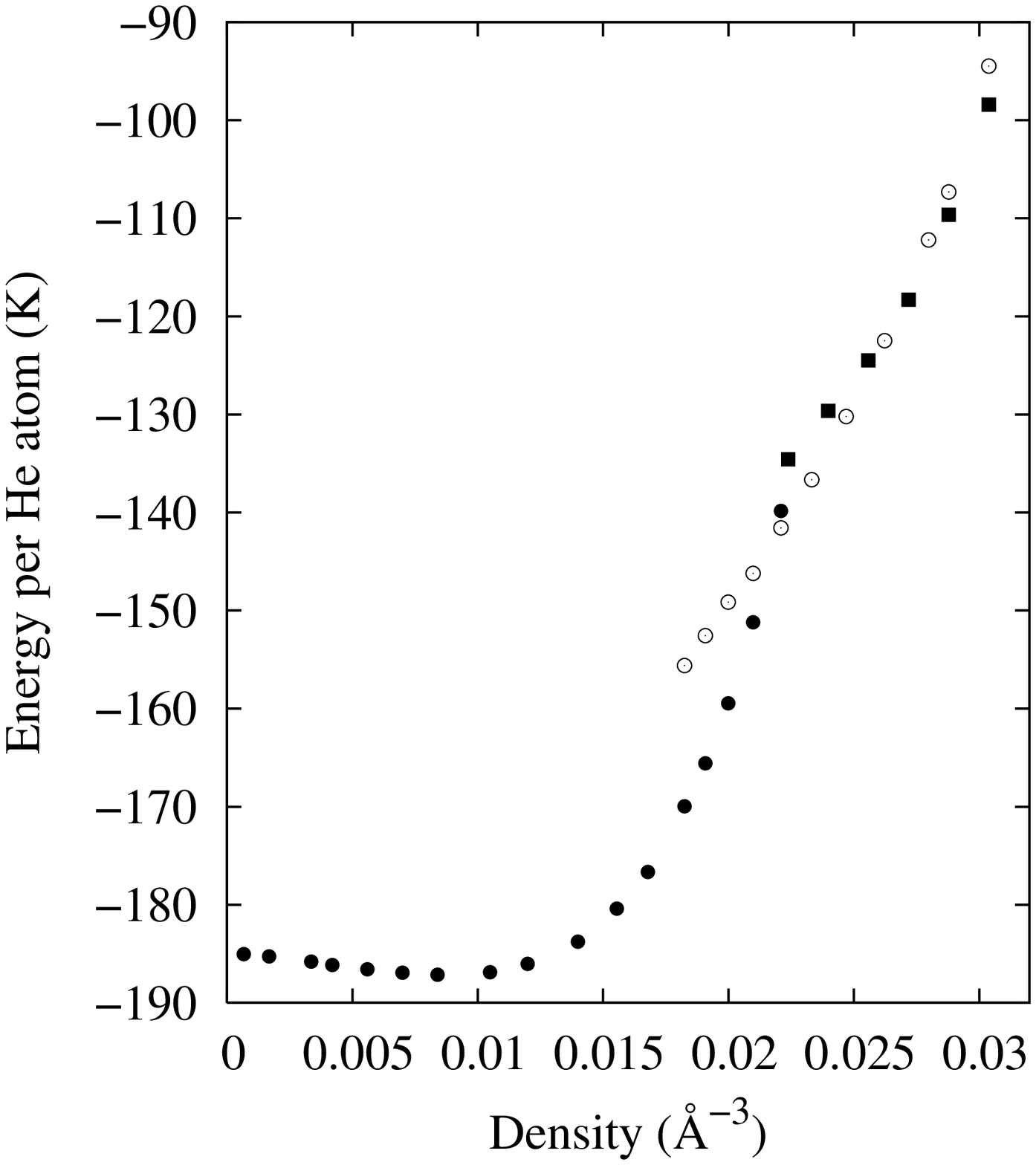}}
\caption{
Energy versus density for different phases: One layer liquid (full circles); 
(squares); two layer liquid (circles); a single layer solid close to the wall 
(full squares). 
}
\label{fig8}
\end{figure}

In order to establish rigorously 
which phases are stable, a double-tangent Maxwell construction      
should be made. In practice, this means to display the free energies 
(energies in the 0 K limit) per atom for each
considered phase  (single layer, and double layer liquid, single layer
solid and single layer solid plus a liquid) versus the inverse of the 
density and draw a single line that connects zones with the same slope. 
Since the slope is minus the equilibrium pressure, if there are several 
possibilities, the line with the minimum slope should be chosen. The 
phases connected by that line will be the stable ones, and the equilibrium
densities will be the ones that share the same slope with the line drawn. 

\begin{figure}
\centering {\includegraphics[width=0.7\linewidth]{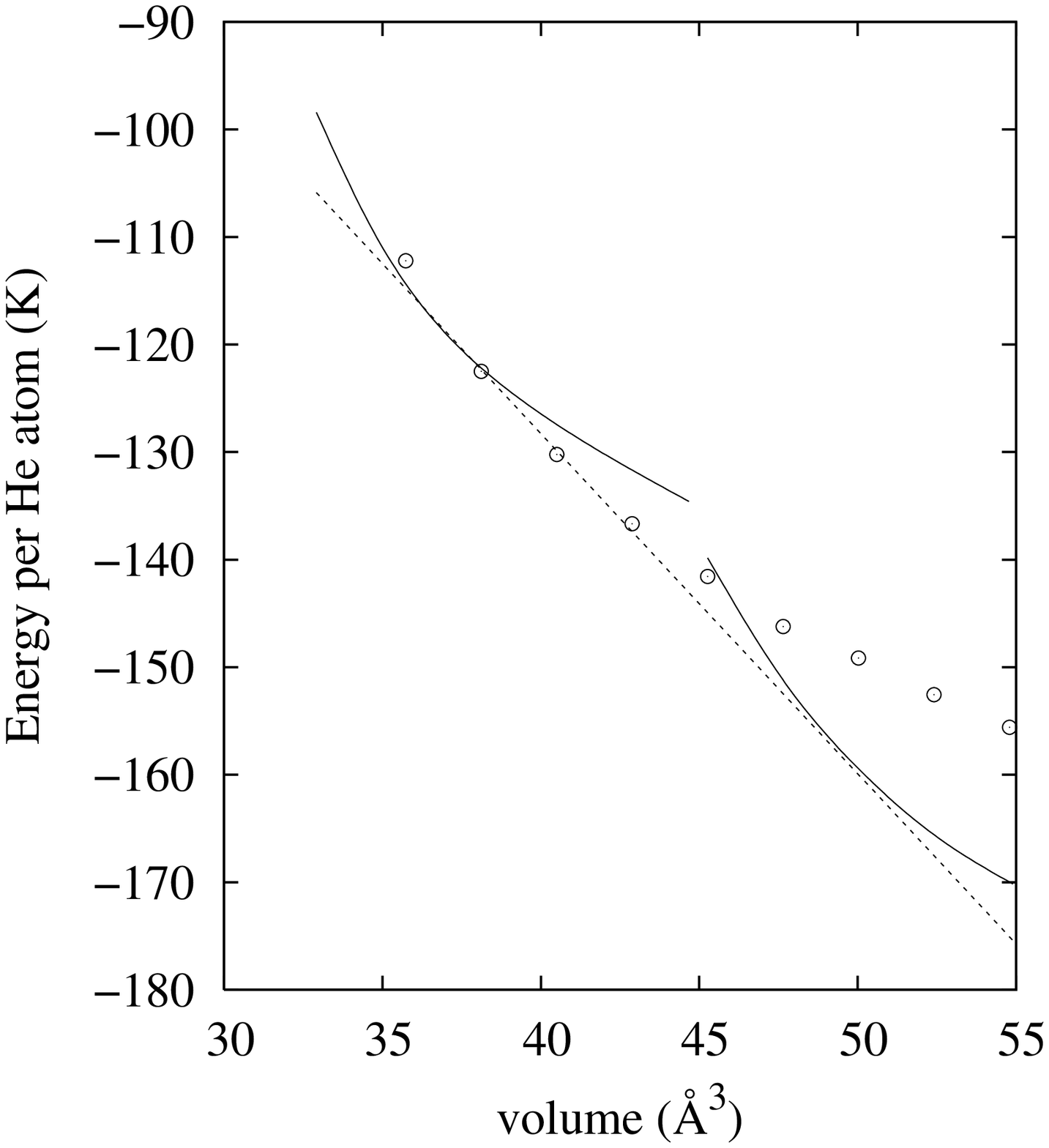}}
\caption{
Maxwell construction to determine the most stable phase inside a (10,10) nanotube
in the density range considered.  
Full line, one layer liquid (lower) and solid+liquid phases (upper); 
Circles, two layer liquid phase.  
}
\label{fig9}
\end{figure}

This procedure is shown in Fig. 9 for the (10,10) tube: the dashed line is a double tangent 
construction between the single-layer liquid and the single-layer 
solid with a liquid on top. This line implies a transition between a 
low density phase of 2 10$^{-2}$ \AA$^{-3}$ and another  
of $2.7$ 10$^{-2}$ \AA$^{-3}$. The transition 
pressure would be of 430 atm. However, the data cannot allow us to 
distinguish between this alternative and to draw  
two transition lines, one between
a single-layer liquid and a double-layer liquid, followed by other
between this last dense liquid and the solid with a liquid on top. 
In this scenario, in addition to the already mentioned equilibrium 
densities, we will have small a window with a two-layer 
liquid at around 2.5 10$^{-2}$ \AA$^{-3}$. 
The one-shell liquid--two-shell liquid transition pressure would be of 420 atm. 
Our results show that the high-pressure stable phase in the density range 
considered is a solid with a liquid on top. This one-layer shell liquid
in a cylindrical arrangement is unique, since simulations of helium 
in wider Gelsil pores indicate that the first layer is a 2D solid 
\cite{reatto}.

As in the previous analysis of the (5,5) tube, it is relevant to take 
into account the influence of the other nanotubes within a bundle arrangement. 
Now, for the (10,10) tubes each 
cylinder is surrounded by six others, with a minimum distance between
centers of  17 \AA \cite{Tersoff}.  The influence that the surrounding tubes 
have in the energy per helium atom
in a particular cylinder can be estimated by the mean-field expression 
\begin{eqnarray}
\lefteqn{E_{correction}  =  6 \int_{x'} \int_{y'} d(x',y') dx' dy'}  \\
& & \times \int_0^\infty \int_x \int_y R(x,y,z) V(x,y,z,x',y') dx dy
dz  \ ,
\nonumber
\end{eqnarray}
where $d(x',y')$ represents the normalized probability of finding an $^4$He
atom at coordinates $x'$ and $y'$ for any $z$ position in the first tube.   
$R(x,y,z)$ is the radial density function
of helium atoms in the second 
tube ($x,y,z$) and $V(x,y,z,x',y')$ is the helium-helium Aziz potential
used in our simulations \cite{aziz}. This integral has been performed 
for all the helium densities corresponding to stable phases  
and verified that its net effect was to decrease the energy per atom, 
but in negligible amounts. For instance, in the solid + liquid  
phase, the energy per atom in an isolated tube was -98.41 $\pm$ 0.09 K, 
with an effect due to the other tubes of -0.11 K. For lower helium 
densities the effect was even smaller.    

Unfortunately, all this theoretical effort cannot be trusted at present by reliable 
experimental data.  One of the main reasons for explaining the difficulties of experiment 
is that for helium to enter inside a tube this has to be opened. However, 
there is a place in which quantum gases are readily adsorbed, i.e., the 
external surfaces of a bundle of carbon nanotubes. In this case there are a number of
experimental measures \cite{teizer2, teizer, vilchesjltp1,
vilchesjltp2,vilcheshe1,
vilcheshe2,lasjaunias,glyde,vilchesne,talapatra2} one can compare the
theoretical results with. Among the theoretical studies of this particular adsorption,
we would mention the case of a single atom or a small cluster in a groove \cite{siber,kro}
and a full DMC calculation on the same subject \cite{yo4,yo5}. This 
last calculation analyzed all the possible phases of $^4$He on  
the outer surface of a (10,10) carbon-nanotube bundle, from a pure
1D one in between two cylinders (what is termed a groove), to a
two layers shell in which the helium layer closer to the surface of
the tubes is a quasi-two-dimensional solid.   

\begin{figure}
\centering {\includegraphics[width=0.7\linewidth]{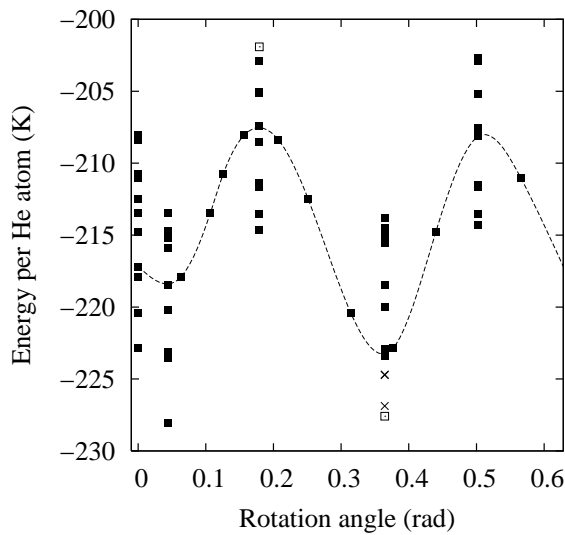}}
\caption{
Energy per helium atom on top of a groove in the infinity dilution limit 
(full squares).  
A dashed line joints the results obtaining by fixing the orientation
of a cylinder and rotating the other one. Additional rotations and 
displacements in the $z$ direction (crosses) are considered. 
Open squares represent the maximum and minimum values of $^4$He 
binding energies. 
}
\label{fig10}
\end{figure}

Fig. 10 displays the binding energy of a single $^4$He atom on the
groove in between two (10,10) carbon nanotubes. Since all the 
C-He interactions are individually taken into account, that energy 
depends on the corrugation of the substance, in particular of the 
relative orientation of the two cylinders that form the groove. The
additional  trial function apart from the product      
of Eq. 12 and  14 is that of Eq. 17 with $b_s$ = 0. $x_{site}$ and
$y_{site}$ are here the optimized coordinates for a single atom left to
roam freely along the length of the groove ($z$ coordinate).    

The absolute maximum and minimum helium energies at 0 K considering all 
rotations and translations of both cylinders  are 
indicated by two open squares and correspond to binding energies of 
227.54 $\pm$ 0.01 and 201.85 $\pm$ 0.01 K, respectively.
These compare favorably with the experimental results given in Ref. 
\onlinecite{teizer2} (range between 210 and 250 K),
and Ref. \onlinecite{vilcheshe2}  (212 K, to be compared to our average 
of $\sim$ 215 K). They also compares favorably with the results given in    
Ref. \onlinecite{kro}, ($\sim$ 211 K). 

The discrepancies of the binding energies for a single atom mean that  
the equation of state of a full array of atoms on top of a groove is going
to be different depending on the particular orientations and displacement
of the tubes forming the groove  with respect to each other. 
To study those cases, 
the same trial function and simulation cell than in the infinite dilution 
limit was used. Fig. 11. shows the results for three different tube
arrangements 
(full squares, maximum binding energy, 227.54 $\pm$ 0.01 K; open squares, 
minimum binding energy, 201.85 $\pm$ 0.01 K; open circles, an intermediate
case). These three calculations are fairly representative  
of all the simulations done: most cases correspond to a quasi-one 
dimensional liquid weakly bounded, while the third case represents the minority
that is a gas. The system is a liquid or a gas depending
on the corrugated structure of the groove, not on the binding energy.
The different relative orientations of the tubes imply also that 
the filling of the grooves will go from the
more binding to the less binding ones up to completion.
This quasi-one dimensional phase has been detected experimentally \cite{lasjaunias, glyde}. 

\begin{figure}
\centering {\includegraphics[width=0.7\linewidth]{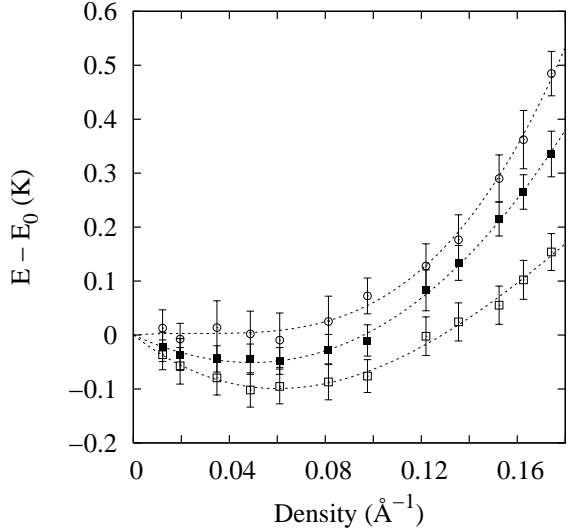}}
\caption{
Energy per helium atom for three representative cases of quasi 1D arrangements. In all cases, 
binding energies for
the infinite dilution limit were subtracted to afford an easier comparison.
}
\label{fig11}
\end{figure}

Fig. 12 shows the next step of the absorption process when more
helium atoms are introduced. In that figure, the $x$ axis is the inverse of 
the surface density. The area of the surface was taken to be as all 
the available space located at $9.5$ \AA \ from the center of any of the 
tubes in the groove. This defined a distance between any adsorbed particle 
and the carbon shell which produced the maximum binding  
energy per particle in all the phases considered. 
In Fig. 12 those phases are the 1D one on a groove (already considered in the
previous figure), what it is called a three-stripes phase (similar
to the previous one but with two other lines of atoms
forming an approximately triangular section) and a 2D liquid
monolayer covering all the external surface of the bundle. 
The corresponding trial function for the three-stripes phase is
the product of the ones given in Eq. 12, 14 and 17 with $b_s$ = 0 and
adequate elections for $x_{\rm site}$,$y_{\rm site}$, while the corresponding
to the 2D liquid substitutes Eq. 17 in the product by
\begin{equation}
\Phi_L = \prod_{i} \exp \left[-a_L (r-r_{\rm center})^2 \right]  \ ,
\end{equation}
where $a_L$ is a parameter variationally optimized and $r$ is the distance
between any of the adsorbed atoms in the liquid phase and the center of
the nanotube. On the other hand, $r_{\rm center}$ = 9.5 \AA \ 
in agreement with the above comment.

\begin{figure}
\centering {\includegraphics[width=0.7\linewidth]{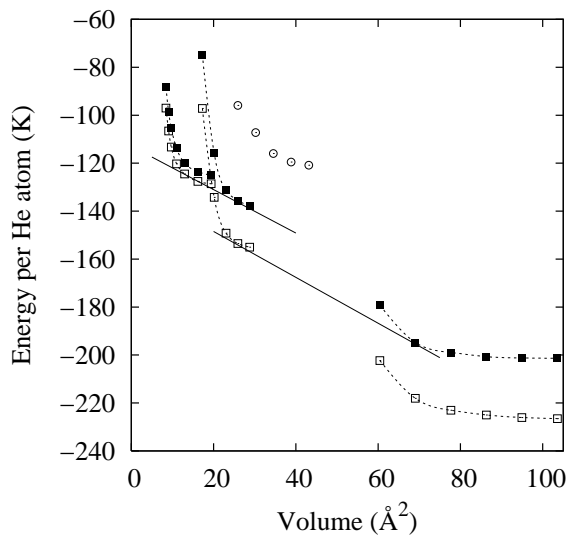}}
\caption{
Maxwell constructions for different helium phases up to the monolayer limit. 
Lower set of curves, quasi 1D system; medium set, 3D phase; 
upper couple of curves, liquid monolayer.  
In all cases, full squares represent the results for the minimum binding 
energies per particle  and open squares for
the maximum ones. Dashed lines are guides-to-the-eye. Open circles 
correspond to an unstable zigzag phase.
}
\label{fig12}
\end{figure}

Since the actual binding energies will depend on the particular configuration
of the groove, only the most extreme cases of minimum and maximum binding
energies per particle were considered for each of the three phases 
already introduced. Within this frame, and from the energy per particle
shown in Fig. 12, one expects the quasi 1D phase to be filled before the 
three stripes (3S) phase starts to do so. That is the reason why the Maxwell 
construction line (lower full line) is made from the
minimum binding energy of the 1D phase to the maximum binding of the 3S one. 
The corresponding equilibrium densities are
$1.4$ 10$^{-2}$ \AA$^{-2}$ (1D) to $3.9$ 10$^{-2}$ \AA$^{-2}$ (three stripes).
This means that the average distance between helium atoms in the 1D phase 
at complete filling is $3.4$ \AA, the same result
found experimentally for a bundle of (8,8) tubes \cite{glyde}. It can also 
be seen that a zigzag phase \cite{gatica2}
(two stripes of helium instead of one or three on top of groove) is
unstable. The upper
full line corresponds to the Maxwell construction between a 2D 
liquid phase and the three stripes one. The corresponding
equilibrium densities are $6.2$ 10$^{-2}$ \AA$^{-2}$ and 
$3.9$ 10$^{-2}$ \AA$^{-2}$. The latter one is the same than for 
the previous transition, indicating a very narrow stability range 
for the 3S phase. This liquid monolayer is akin to the single layer
liquid already described above for helium inside a (10,10) nanotube,
being both the only stable ground-state 2D liquids in contact with a 
carbon monolayer. The other possibility, $^4$He on flat graphene, was found
to have a solid ground state \cite{yo6}. We know that the liquid phase is
the stable one with the help of Fig. 13: the full squares (solid monolayer)
are always on top (higher energies) than other phases with the same 
density.     
  
\begin{figure}
\centering {\includegraphics[width=0.7\linewidth]{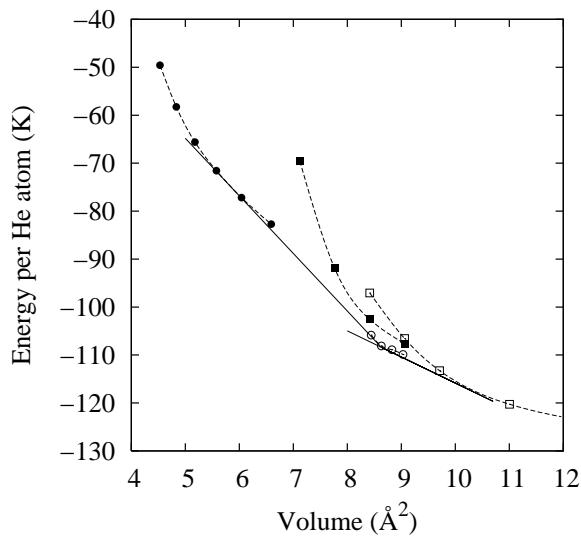}}
\caption{
Phases of helium in the groove at high densities: open squares, 2D liquid monolayer; full squares, 2D solid monolayer;
open circles, 1D phase on top of a liquid; full circles, two layered-phase.  
}
\label{fig13}
\end{figure}

 The results for the minimum
energy structures are the actually shown in Fig. 13. Open circles 
correspond to a quasi 1D phase on top of a liquid monolayer of 
$0.103$ \AA$^{-2}$. Full circles represent the binding energies 
of an structure formed by an eight-channel solid monolayer 
in contact with the carbon shell and density
0.110 \AA$^{-2}$ and a 2D liquid monolayer on top of it located 
at a distance of 12.6 \AA$ $ to the center of closest tube. No other 
structure with lower energy in this density range was
found. Neither of the following possibilities is found:
a double-liquid layer, a three-stripes second-layer phase, and a 
quasi 1D phase on top of a solid layer.    
The second layer quasi 1D phase was found experimentally in the
case of Ne \cite{talapatra2}, but has not been identified for helium. 
There is not any experimental
indication of a solid + liquid phase at 0 K. The experimental results on 
the single monolayer seem to point to a
solid, not to the liquid found here \cite{glyde}. The difference could be due 
to shortcomings of the simulation, (for instance, a too small simulation
cell) or to the fact that the experimental results are for a bundle of 
(8,8) tubes instead of the (10,10) considered here.

\section{Quasi-one-dimensional H$_{\bf 2}$ and D$_{\bf 2}$ in carbon nanotubes}

The physisorption of hydrogen \cite{cole2,dil,dar,rze,wan} in the quest for a fuel 
cell efficient enough to be used as a pollution-free energy carrier has been studied 
in different environments. Single wall carbon nanotubes (SWCN) with diameters of the order 
of a nanometer have been proposed as one of the possible candidates
to approach the pursued level of packing \cite{dil}.   
Besides its technological relevance, the case study of
H$_2$ adsorbed in carbon nanotubes offers the appealing possibility of the
existence of a homogeneous liquid phase at zero temperature. It is worth 
noticing that both, bulk and two-dimensional H$_2$, are solid in the
zero-temperature limit.

Molecular para-hydrogen at zero temperature was studied \cite{prl2000} 
using the DMC method in a 
one-dimensional (1D) array and inside a single walled carbon nanotube (T)
of radius $R$ = 3.42 \AA\  (a (5,5) armchair tube
\cite{ham}) that is one of the narrowest experimentally obtained \cite{iji2}.
H$_2$ molecules interact via the isotropic
semiempirical potential from Silvera and Goldman (SG) \cite{sil1} that has been 
extensively used in path integral Monte Carlo (PIMC) and DMC calculations \cite{chen}. 
The SG is a pair potential that  incorporates to some extent
the effect of  three-body interactions by means of an effective two-body 
term of the form $C_9/r^9$. On the other hand, the isotropy of the potential 
is well justified if one considers that at very low temperatures almost all 
the H$_2$ molecules are para-hydrogen species, i.e., they are in the J=0
rotational state.  In the simulations of H$_2$ inside a nanotube, we use
the cylindrically symmetric potential suggested by Stan and Cole
\cite{cole}. Considering the Lennard-Jones parameters of the pair C-H$_2$
interaction $\sigma$ = 2.97 \AA\ and $\epsilon$ =
42.8 K, the symmetric potential felt by a H$_2$ molecule in a (5,5) tube has a
depth of 42 $\epsilon$, three times larger that the attraction of the same 
molecule in a flat graphitic surface. 

\begin{table}
\begin{center}
\begin{tabular}{lcccc}
$\lambda$ (\AA$^{-1}$) &  $E/N$ (1D, $a$ = 0)  & $E/N$ (1D, $a \not=$ 0)
& $E/N$ (T, $a$ = 0) & $E/N$ (T, $a \not=$ 0)
\\ \hline
0.329  &   98.083 $\pm$    0.034  &   97.963  $\pm$  0.016 &   -1453.99 $\pm$ 0.06 & -1454.69 $\pm$ 0.04 \\
0.320  &   72.567 $\pm$    0.013  &   72.523  $\pm$  0.007 &   -1476.74 $\pm$ 0.05 & -1476.88 $\pm$ 0.01 \\
0.312  &   53.264 $\pm$    0.010  &   53.227  $\pm$  0.010 &   -1493.790 $\pm$ 0.019 &  -1493.720 $\pm$ 0.002 \\
0.304  &   38.581 $\pm$    0.018  &   38.636  $\pm$  0.014 &   -1506.570 $\pm$ 0.03 & -1506.540 $\pm$ 0.011  \\
0.290  &   19.203 $\pm$    0.010  &   19.260  $\pm$  0.003 &   -1523.730 $\pm$ 0.017 & -1523.600 $\pm$ 0.02 \\
\end{tabular}
\caption{Energies per particle in K at high linear densities $\lambda$ for
1D and T H$_2$ systems. $a$ = 0 and $a \not=$ 0 correspond to the liquid
and solid phases, respectively.}
\end{center}
\end{table}

 In Table 3, we show DMC results for the energy per particle  in both 1D and inside a 
(5,5)  nanotube,
and  for the liquid and solid phases.  The comparison between the energies
of both phases at the same density shows that their difference  changes  
sign in going from $\lambda$ = 0.312 \AA$^{-1}$ to $\lambda$ =
0.304 \AA$^{-1}$ in 1D and from $\lambda$ = 0.320 \AA$^{-1}$ to 
$\lambda$ = 0.312 \AA$^{-1}$ in the tube. Above these densities,
 the system prefers to be
localized in a solid-like structure with a difference $|E(s) - E(l)|$ that
increases with $\lambda$. When the density decreases 
the liquid phase is energetically preferred and again the
size of the difference $|E(s) - E(l)|$ increases when $\lambda$ diminishes. 
The density value at which this difference becomes zero is estimated to be 
$\lambda = 0.309$ \AA$^{-1}$ in 1D and $\lambda = 0.315$ \AA$^{-1}$ in the
tube, being not possible to distinguish between 
freezing and melting densities.

Inside the nanotube (T), the energies are much more negative that in 1D due
to the strong attraction of the carbon substrate: the binding energy of a
single H$_2$ molecule in the tube is $E_b = -1539.87 \pm  0.11$ K.  Looking at
the T-energy results contained in Table 3 one realizes that also in this 
case a transition occurs at a density very close to the
1D one. It is remarkable that both in 1D and T, H$_2$  remains liquid 
below the liquid-solid transition density, and thus a homogeneous liquid
phase at zero pressure is predicted. That result contrasts with the
theoretically and experimentally well established solid phase in 3D
\cite{sil2} and the
2D solid phase predicted by a PIMC calculation \cite{wag}.

\begin{figure}
\centering {\includegraphics[width=0.7\linewidth]{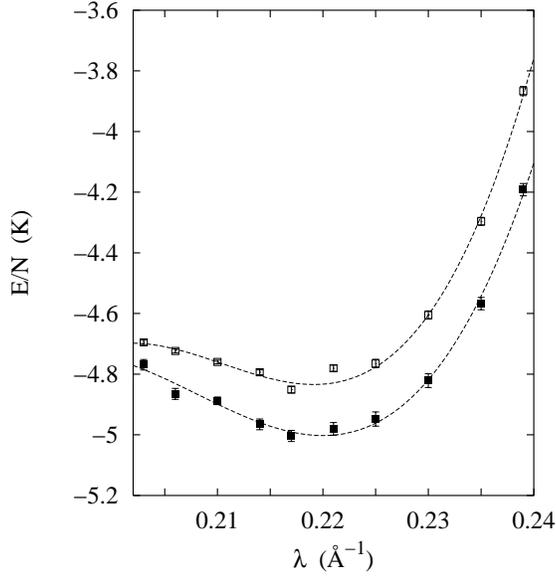}}
\caption{Energy per particle of H$_2$ as a function of the linear density. Open
squares are the 1D results, and filled squares are the T energies having
subtracted the binding energy of a single molecule $E_b$. The lines are the
result of polynomial fits to the DMC data.}       
\end{figure} 
  
The equations of state of liquid H$_2$ near the equilibrium density for 
both the 1D and T systems are shown in Fig. 14. In order to make the energy
scales compatible we have subtracted the single binding energy $E_b$ to the
T results. The lines in the figure correspond to third-degree polynomial
fits.  The equilibrium densities in both systems are the same considering
their respective uncertainties but the binding energy $e_0 = e(\lambda_0)$
is larger when H$_2$ is inside the nanotube.

In Fig. 15, the density dependence of the pressure for both the 1D and T 
systems is reported from equilibrium up to the liquid-solid transition
density. Results  for $^4$He are also
plotted for comparison. Both in H$_2$ and  $^4$He the pressure increases faster in the
1D geometry ($P_{\lambda}$) than in the tube (P) due the transverse degree of
freedom that particles have in the latter case (notice the proportionality
between the scales of $P$ and $P_{\lambda}$ in Fig. 2, $P_{\lambda}/P = \pi
R^2$). At a given density $\lambda$, the difference between the T and 1D
pressures is smaller in H$_2$ than in $^4$He. 
The one-dimensionality of H$_2$ inside the nanotube
is observed in all the liquid regime in contrast with  
$^4$He, in which the departure from such an idealized model already appears
around the equilibrium density and increases significantly with $\lambda$ (see Section 3).  
Also apparent from Fig. 15 is a much smaller compressibility in H$_2$ than
in $^4$He.

\begin{figure}
\centering {\includegraphics[width=0.7\linewidth]{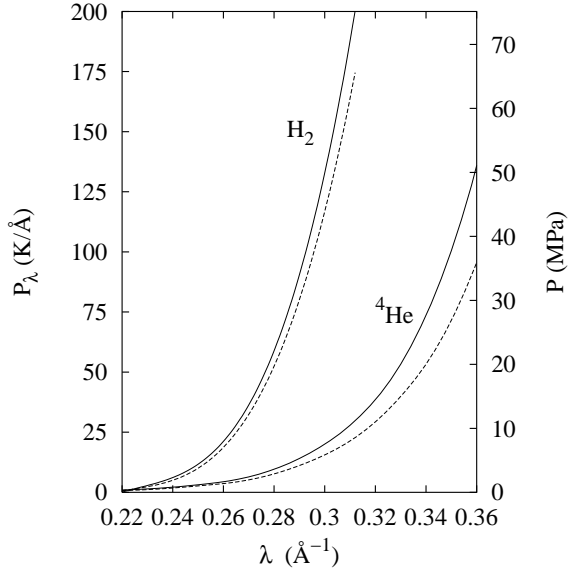}}
\caption{1D ($P_{\lambda}$, solid line) and T ($P$, dashed line)  
pressures for H$_2$ and $^4$He as a function of the linear density.}                                
\end{figure}    

The DMC method has also been  used to study the influence of 
both the interparticle potential and mass 
on the thermodynamic behavior of the isotopes of molecular hydrogen, 
H$_2$ and D$_2$, adsorbed inside a carbon nanotube \cite{deu}.
Since the electronic structure of molecular deuterium and hydrogen 
is the same, the D$_2$-D$_2$ interparticle potential is
identical to the H$_2$-H$_2$ one. This is equally true 
for the particle-tube interactions. Moreover, the mass of the 
D$_2$ molecule is very similar to that of a $^4$He atom. Thus, by
comparing the deuterium results with those for helium, 
the effect of the respective potential wells can be inferred. On the
other hand, the influence of the 
zero-point energy in the thermodynamic behavior of a quasi-one 
dimensional array can be drawn from the comparison between the deuterium and
hydrogen results. 

At low densities,
the dominant effect is due to the binding energy of a single 
molecule to the nanotube.  From our DMC calculations the binding energy of a single 
molecule inside a carbon tube turns out to be, for
the (5,5) tube,
$-1539.87 \pm 0.11$ K for H$_2$ , and $-1605.23 \pm 0.01$ K and $-1624.37 \pm
0.01$ K for 
D$_2$ and T$_2$, respectively. The increase in the binding energy with the mass
comes from the combination of two features: a decrease in the kinetic
energy, mainly due to a direct effect of the mass ($m_{{\rm H}_2}/m _{{\rm
D}_2} \simeq 1/2$, $m_{{\rm D}_2}/m _{{\rm T}_2} \simeq 2/3$), and a
simultaneous increase of the potential energy.
Although those
energies correspond to the ground state of single molecules at 0 K,
they constitute a very good estimation in the limit of infinite dilution
at nonzero temperatures.
From the above binding energies one can extract information on the
selectivity in the adsorption inside the nanotube.
Following Ref. \onlinecite{sie}, the selectivity of isotope 2 with 
respect to isotope 1 
can be defined by the quotient
$S=(x_1/x_2)/(y_1/y_2)$ with $x_i$ ($y_i$) the nanotube (bulk) mole
fractions. It has been proved that in the limit of zero pressure the selectivity
$S_0$ is very well approximated by
\begin{equation}
S_0 = \frac {m_2}{m_1} \ \exp \left( - \frac{E_{1}-E_{2}}{T} \right) \ ,
\label{select}
\end{equation}
where $E_i$ is the binding energy of isotope $i$. 
Considering $T=20$ K, as in Ref. \onlinecite{sie}, we obtain 
$S_0({\rm T}_2 / {\rm H}_2 ) = 22.8$ and $S_0({\rm T}_2 / {\rm D}_2 ) = 1.7$
for the (5,5) tube.  The selectivity is especially high in
the case T$_2$/H$_2$ due to the sizeable difference in binding energies
between the two isotopes, $E_{{\rm T}_2} - E_{{\rm H}_2} = -84.5$ K. That
large selectivity, which is a purely quantum effect, has been proposed in
Ref. \onlinecite{sie} to achieve an efficient isotopic sieving.

\begin{table}
\begin{center}
\begin{tabular}{lcc}
          &  1D D$_2$  & D$_2$ in a tube \\ \hline
$\lambda_0$ (\AA$^{-1})$ &  0.2457 $\pm$ 0.0003 
& 0.2473 $\pm$ 0.0002 \\
$(E/N)_0$ (K)                 
& -10.622 $\pm$ 0.016 & -1615.94 $\pm$ 0.015 \\
$A$   (K)              & 2.0 10$^2$  
$\pm$ 1.0 10$^1$ & 2.13 10$^2$ $\pm$ 1.0 10$^1$ \\
$B$   (K)                 
& 9.6 10$^2$  $\pm$ 1.2 10$^2$ & 1.10 10$^3$ $\pm$ 1.1 10$^2$ \\
\end{tabular}
\caption{Parameters of the equation of state of D$_2$.}
\end{center}
\end{table}

\begin{figure}[b]
\centering {\includegraphics[width=0.7\linewidth]{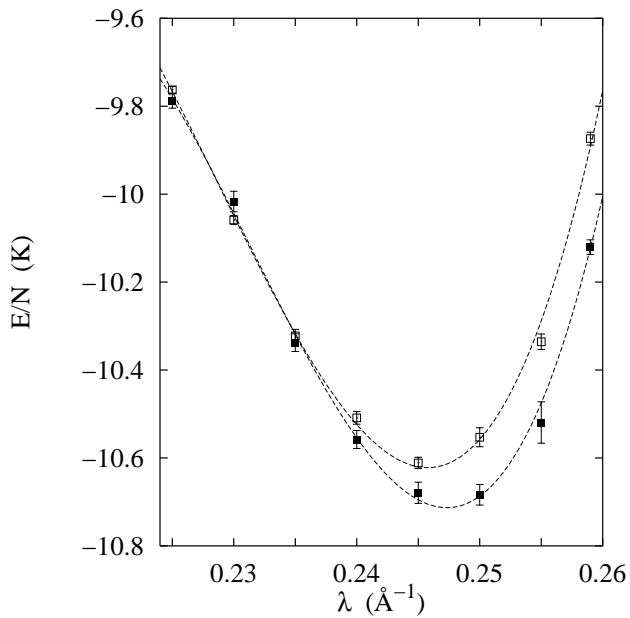}}
\caption{Comparison between the equation of state of  1D D$_2$ and D$_2$
adsorbed in the nanotube. Filled squares correspond to the tube results;
open squares, to the 1D ones. The lines are polynomial fits to the DMC
data. To better compare both results, we have subtracted to the tube
energies the binding energy of a single molecule.} 
\end{figure}

DMC energy results for both 
1D D$_2$ and D$_2$ adsorbed in the (5,5) nanotube
are displayed in Fig. 16.
In order to show the two equations of state with the same energy
scale, we have subtracted to the tube results the adsorption 
energy of a single molecule.
In Fig. 16, the curves are  polynomial fits to the DMC data,
the optimal parameters for the tube being reported in Table 4. It 
can be seen that the equilibrium density for 1D D$_2$ and D$_2$ adsorbed 
in the (5,5) nanotube are almost identical. This is also true for
the location of the spinodal points of D$_2$, 
$\lambda_{\rm s}^{\rm 1D} = 0.230 \pm
0.001$ \AA$^{-1}$ and $\lambda_{\rm s}^{\rm T} = 0.232 \pm 0.001$
\AA$^{-1}$,
which can be derived from the data contained in Table 4.

\begin{figure}
\centering {\includegraphics[width=0.7\linewidth]{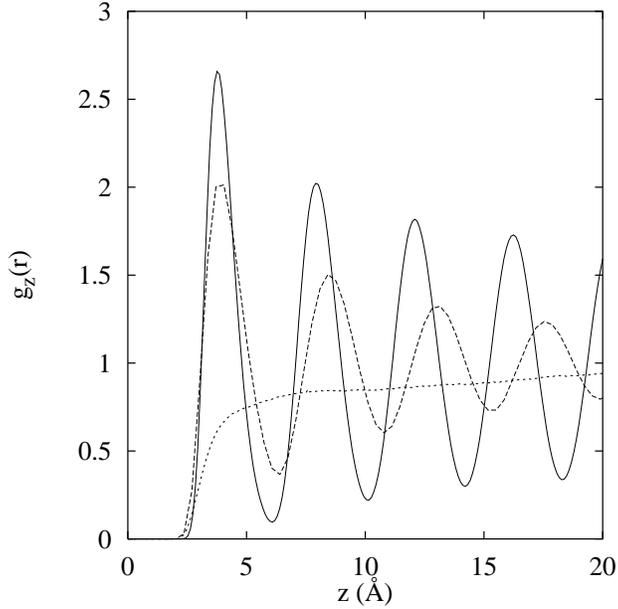}}
\caption{Two-body distribution function in the nanotube system, and 
in the $z$ direction. Solid,
dashed, and dotted lines correspond to D$_2$, H$_2$, and $^4$He,
respectively. All curves are calculated at their respective equilibrium
densities $\lambda_0$.}
\end{figure}

The energies in the respective equilibrium points ($(E/N)_0$) 
are slightly different: the additional transverse degree of freedom 
only amounts to an increase of 0.091 K. This increase in the binding energy 
is nearly a factor two smaller than the one drawn from the DMC calculations 
for H$_2$ (0.172 K). In relative terms, the increase of the binding energy 
is only a 0.85 \% for D$_2$ versus a 3.5 \% for H$_2$.
Therefore, the effects of the additional degree of 
freedom of the D$_2$ molecules in the radial direction inside the nanotube, 
which account for the enhancement of the binding energy, are reduced by 
the greater mass of the D$_2$ molecule with respect
to the H$_2$ one. As a matter of comparison, it is illustrative to compare
the effects observed in D$_2$ with the ones previously studied in $^4$He
using the same methodology and geometry. It is worth noticing that the
masses of D$_2$ and $^4$He are nearly the same whereas the interatomic
potentials are sizably different. The DMC results show that the latter effect
is completely determinant: in  $^4$He the relative difference mentioned 
above is 90 \%, two orders of magnitude 
larger than in D$_2$. Another minor effect that contributes to the 
one-dimensionality of molecular deuterium adsorbed inside the tube, 
is the larger hard-core size of the 
C-D$_2$ interaction ($\sigma_{{\rm C}-{\rm D}_2} = 2.97$ \AA), versus the
C-He one
($\sigma_{{\rm C}-{\rm He}} = 2.74$ \AA). 
The mass versus potential effects can also be seen in the value of the
equilibrium density. Inside the tube, $\lambda_0$ goes from $0.079 \pm 0.003$
\AA$^{-1}$ in $^4$He, to $0.2200 \pm 0.0006$ \AA$^{-1}$ in H$_2$, to reach
$0.2473 \pm 0.0002$ \AA$^{-1}$ in D$_2$. That sequence clearly indicates
that the main 
influence in $\lambda_0$ comes from the potential, since the isotopic change 
varies the
location of the energy minimum less than 15 \%. The features 
observed in these systems have probably a more general character and one can guess
that if the well of the interatomic potential is increased 
and/or the mass of the particle adsorbed inside a tube is enlarged, the effect
would be an increase in the $\lambda_0$ value.

The microscopic study of the spatial structure of the molecules in the array
provides additional and useful information on the system.
In Fig. 17, results for the radial distribution functions, $g_z(r)$, along the $z$ axis, 
are shown.  They correspond to the quantum fluids adsorbed in the tube at
their respective equilibrium densities $\lambda_0$. 
Being the denser of the three systems, D$_2$ exhibits accordingly the
most pronounced oscillations in the $g_z(r)$ function.
The shift in the positions of the maxima for the two molecular isotopes
arises basically from the difference in their respective $\lambda_0$'s.  
In the $^4$He case, the much lower 
equilibrium density, which is direct consequence of the different potential,
explains the smoothness of the  $g_z(r)$ obtained.

\begin{figure}
\centering {\includegraphics[width=0.7\linewidth]{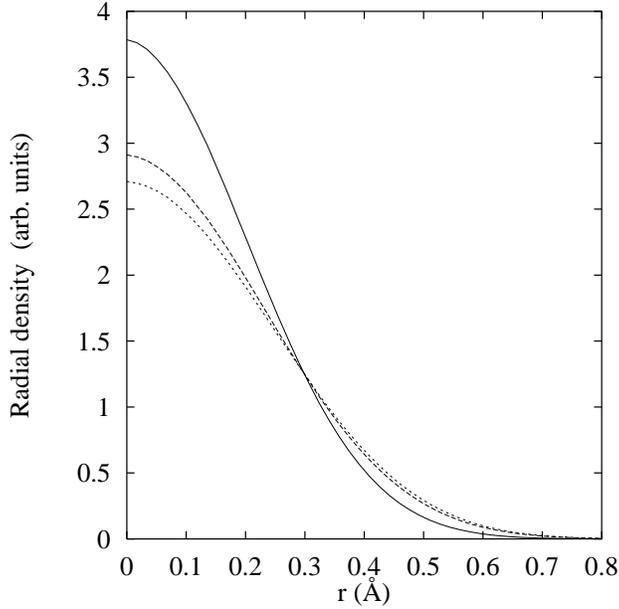}}
\caption{Radial density of D$_2$ (solid line), H$_2$ (dashed line), and
$^4$He (dotted line) inside the (5,5) nanotube.}
\end{figure}

The radial densities inside the (5,5) tube have  been also studied.
In Fig. 18,  the radial densities for $^4$He, H$_2$, and D$_2$
for the same linear density $\lambda$ = 0.245 \AA$^{-1}$ are shown. 
The trends shown in the figure are common to all densities studied:
the particle 
with the largest  mass (D$_2$) is the one which spends more time in regions 
closer to the center of the tube, i.e., D$_2$ in the tube is the closest to a
one-dimensional system. The change in the mass  and in the
interatomic potential  work in the same direction: the radial
densities of H$_2$ and $^4$He are quite similar. Both curves show a
decrease in the radial localization and larger fluctuations
in the transverse direction.

\section{Concluding remarks}

We have reviewed work carried out mainly in our group on the microscopic
description of quantum liquids adsorbed in single nanotubes or in a bundle
of them. This theoretical approach has been made by using quantum Monte
Carlo, mainly the diffusion Monte Carlo method that allows for a very
accurate study of the ground state of the system starting on basic
knowledge: the mass, the geometry of the environment and the
interatomic potentials. The results here presented correspond to the fully
quantum fluids $^4$He, H$_2$, and  D$_2$. Apart from some technological
applications to come, mainly for hydrogen, the study of quantum fluids and
solids in nanotubes offer the real possibility of having nearly
one-dimensional quantum systems, extending the previous well-known
confining geometry of fluids adsorbed on planar surfaces like graphite
which provides a quasi-two-dimensional environment.

A bundle of nanotubes presents the interesting feature of different
adsorption places that deserve particularized attention: the inner part of a
nanotube, the intersite channels between three neighboring tubes, and the
grooves formed in the external surface of the bundle. In the case of $^4$He
we have reported results on these different sites taking also into account
the effects induced by other atoms filling neighboring tubes. When the
atoms or molecules are adsorbed in the inner part of a single tube one
observes a quasi-one dimensional system, mainly when the tube is very
narrow. The comparison between the equations of state and structural
properties of the fluid inside the tube and in a purely 1D geometry allows
for a meaningful estimation of the proximity between both systems. In the
same narrow tube, we have verified that H$_2$, and mainly D$_2$, approaches
better to the 1D geometry than $^4$He due to their stronger interaction
with respect to helium. Importantly, our DMC results proved that the ground
state of para-H$_2$ inside a (5,5) nanotube is a liquid in contrast with
their well-established solid phases in bulk and in 2D.

The experimental confirmation of the results presented in this review is
still pending to a big extent. After some preliminary interpretation of
experiments carried out with different species about possible adsorption in
the inner part of nanotubes or even in the intersites, a more careful
analysis concluded the practical impossibility  of the actual determination 
of the dominant adsorption surfaces. Now, it is more generally assumed that
the filling of the tubes is quite improbable and that the gases in contact
with a nanotube bundle are predominantly adsorbed on the grooves and on the
external surface \cite{vilchesjltp1,vilchesjltp2}. A possible mechanism that could explain the limited
uptake of the tubes is the present of defects in the carbon lattice. We
have presented explicit calculations of the influence of this vacancy and
proved that this can effectively reduce the adsorption inside a nanotube.

\begin{acknowledgements}
We acknowledge financial support from DGI (Spain) Grants No. FIS2006-02356
and  FIS2008-04403, Junta de Andalucia Grant No. FQM-205, and
Generalitat de Catalunya Grant No. 2005SGR-00779.
\end{acknowledgements}

\pagebreak


\begin{thebibliography}{99}

\bibitem{Iijima} S. Iijima, Nature (London) {\bf 354}, 56 (1991). 

\bibitem{harris} P.J. Harris, {\it Carbon Nanotubes and Related Structures} 
(Cambridge University Press, Cambridge UK, 1999).

\bibitem{reich} S. Reich, C. Thompsen, and J. Maultzsch, {\it Carbon
Nanotubes} (Wiley, Berlin, 2004). 

\bibitem{pro1} C. Journat and P. Bernier, Appl. Phys. A. {\bf 67}, 1
(1998).

\bibitem{pro2} M.J. Bronikowki, P.A. Willis, D.T. Colbert, K.A. 
Smith, and R.E. Smalley, J. Vac. Sec. Technol. A {\bf 19}, 
1800 (2001).  

\bibitem{pro3} M.S. Dresselhaus, G. Dresselhaus, and P. Avouris, 
{\it Carbon Nanotubes}, (Springer, Berlin, 2001). 

\bibitem{coleRMP} M.M. Calbi, M.W. Cole, S.M. Gatica, M.J. Bojan, and
G. Stan, Rev. Mod. Phys. {\bf 73}, 857 (2001).

\bibitem{boro94} J. Boronat and J. Casulleras, Phys. Rev. B {\bf 49},
8920 (1994). 

\bibitem{aziz} R.A. Aziz, F.R.W McCourt, and C.C. K. Wong, Mol. Phys.
{\bf 61}, 1487 (1987). 

\bibitem{silvera} I.F. Silvera and V.V. Goldman, J. Chem. Phys. 
{\bf 69}, 4209 (1978). 

\bibitem{cole} G. Stan and M.W. Cole, Surf. Sci. {\bf 395}, 280 (1998).

\bibitem{cole2}  G. Stan and M. W. Cole, J. Low. Temp. Phys. {\bf 110}, 539 
(1998).

\bibitem{uptake} G. Stan, M.J. Bojan, S. Curtarolo, S.M. Gatica, and M.W. Cole,
Phys. Rev. B, {\bf 62}, 2173 (2000).

\bibitem{he4}  M.C. Gordillo, J. Boronat, and J. Casulleras, Phys. Rev. B {\bf 
61}, R878 (2000). 

\bibitem{azizold} R. A. Aziz, V. P. S. Nain, J. S. Carley,  W. L. Taylor,
and G. T. McConville, J. Chem. Phys. {\bf 70}, 4330 (1979).

\bibitem{moroni} M. Boninsegni and S. Moroni, J. Low. Temp. Phys. {\bf
118}, 1 (2000).

\bibitem{krot} E. Krotscheck and M. D. Miller, Phys. Rev. B {\bf 60},
13038 (1999).  

\bibitem{prl2000} M.C. Gordillo, J. Boronat, and J. Casulleras, Phys. Rev. Lett.
{\bf 85}, 2348 (2000)

\bibitem{deu}  M.C. Gordillo, J. Boronat, and J. Casulleras, 
Phys. Rev. B {\bf 65}, 014503 (2002).

\bibitem{llorens1}  L. Brualla and M.C. Gordillo, Phys. Rev. B  {\bf 68}, 075423
(2003).  

\bibitem{pro} M.C. Gordillo, J. Boronat, and J. Casulleras
in {\em Proceedings of the 14th international conference "Recent
progress in many-body theories"} Series on Advances in Quantum Many Body Theory.
Vol 11. Ed. J. Boronat, G.E. Astrakharchick, and F. Mazzanti, (World
Scientific, Singapore, 2008). 

\bibitem{teizer2} W. Teizer, R.B. Hallock, E. Dujardin, and T.W. Ebbesen, 
Phys. Rev. Lett. {\bf 84}, 1844 (2000).

\bibitem{cpl} D.B. Mawhinney, V. Naumenko, A. Kuznetsova, J.T. Tates Jr.,
J. Liu, and R.E. Smaley, Chem. Phys. Lett. {\bf 324}, 213 (2000).

\bibitem{adv} E. Dujardin, T.W. Ebbesen, A. Krishnan, and M.M.J. Treacy,
Adv. Mat. {\bf 10}, 611 (1998). 

\bibitem{lu} A.J. Lu and B.C. Pan, Phys. Rev. Lett. {\bf 92}, 105504 (2004).

\bibitem{ajayan} P.M. Ajayan, V. Ravikumar, and J.C. Charlier, Phys. Rev. 
Lett. {\bf 81}, 1437 (1998). 

\bibitem{kra} A.V. Krasheninnikov, K. Nordlund, 
M. Sirvi\"o, E. Salonen, and J. Keinonen,
Phys. Rev. B {\bf 63}, 245405 (2001). 

\bibitem{zhang} S. Zhang, S.L. Mielke, R. Khare, D. Troya, R.S. Ruoff, 
G.C. Schatz, and T. Belytschko, Phys. Rev. B {\bf 71}, 115403 (2005).

\bibitem{prl06} M.C. Gordillo, Phys. Rev. Lett. {\bf 96}, 216102 (2006).

\bibitem{yoh2} M.C. Gordillo, Phy. Rev. B {\bf 76}, 115402 (2007).

\bibitem{bony} M. Boninsegni, S. Y. Lee, and V. H. Crespi,
 Phys. Rev. Lett. {\bf 86}, 3360 (2001). 

\bibitem{low}  J. Boronat, M.C. Gordillo, and J. Casulleras, 
J. Low Temp. Phys. {\bf 126} 199 (2002).

\bibitem{vilcheslt25} O.E. Vilches, communication to LT25. 

\bibitem{colemf}  M. W. Cole, V. H. Crespi, G. Stan, C. Ebner, J. M.
Hartman, S. Moroni, and M. Boninsegni, Phys. Rev. Lett. {\bf 84}, 3883
(2000).

\bibitem{interh2} M.C. Gordillo, J. Boronat, and J. Casulleras. Phys. Rev. B
{\bf 68}, 125421 (2003). 

\bibitem{llorens2} M.C. Gordillo, L. Brualla, and S. Fantoni.  Phys. Rev. B 
{\bf 70}, 245420 (2004).

\bibitem{yo3} M.C. Gordillo, J. Boronat, and J. Casulleras.  Phys. Rev. B 
{\bf 76}, 193402 (2007). 

\bibitem{reatto} M. Rossi, D.E. Galli, and L. Reatto. Phys. Rev. B. {\bf
72},
064516 (2005).  

\bibitem{Tersoff} J. Tersoff and R.S. Ruoff. Phys. Rev. Lett. {\bf 73}, 676
(1994).

\bibitem{teizer} W. Teizer, R.B. Hallock, E. Dujardin, and T.W. Ebbesen.
Phys. Rev. Lett. {\bf 82}, 5305 (1999).

\bibitem{vilchesjltp1} T. Wilson, A. Tyburski, M.R. DePies, O.E. Vilches,
D. Becquet, and M. Bienfait, J. Low Temp. Phys. {\bf 126}, 403 (2002).

\bibitem{vilchesjltp2} S. Ramachandran, T.A. Wilson, D. Vandervelde,
D.K. Holmes, and O.E. Vilches, J. Low Temp. Phys. {\bf 134}, 115 (2004).

\bibitem{vilcheshe1} Y.H. Kahng, R.B. Hallock, E. Dujardin, and T.W.
Ebbesen,
J. Low. Temp. Phys. {\bf 126}, 223 (2002).

\bibitem{vilcheshe2} T. Wilson and O.E. Vilches, Physica B {\bf 329-333},
278 (2003).

\bibitem{lasjaunias} J.C. Lasjaunias, K. Biljakovic, J.L. Sauvajol,
and P. Monceau, Phys. Rev. Lett. {\bf 91}, 025901 (2003).

\bibitem{glyde} J.V. Pearce, M.A. Adams, O.E. Vilches, M.R. Johnson, and
H.R. Glyde, Phys. Rev. Lett. {\bf 95}, 185302 (2005).

\bibitem{vilchesne} S. Ramachandran and O.E. Vilches, Phys. Rev. B.
{\bf 76}, 075404 (2007).

\bibitem{talapatra2} S. Talapatra, V. Krungleviciute, and A.D. Migone,
Phys. Rev. Lett. {\bf 89}, 246106 (2002).

\bibitem{siber} A. Siber, Phys. Rev. B. {\bf 66}, 205406 (2002).

\bibitem{kro} M. Aichinger, S. Kili{\'c}, E. Krotscheck, and L. Vranje\v{s},
Phys. Rev. B {\bf 70}, 155412 (2004).

\bibitem{yo4} M.C. Gordillo, Phys. Rev. Lett. {\bf 101}, 046102 (2008).

\bibitem{yo5} M.C. Gordillo, J. Phys.: Conf. Ser. {\bf 150}, 032023 (2009).

\bibitem{gatica2} M.M. Calbi, S.M. Gatica, M.J. Bojan, and M.W. Cole,
J. Chem. Phys. {\bf 115}, 9975 (2001).

\bibitem{yo6} M.C. Gordillo and J. Boronat, Phys. Rev. Lett. {\bf 102},
085303 (2009).  

\bibitem{dil} A. C. Dillon, K. M. Jones, T. A. Bekkedahl, C. H. Kiang, D.
S. Bethune, and M. J. Heben, Nature {\bf 386}, 377 (1997).

\bibitem{dar} F. Darkrim and D. Levesque, J. Chem. Phys. {\bf 109}, 4981
(1998).

\bibitem{rze} M. Rzepka, P. Lamp, M. A. de la Casa-Lillo, J. Phys. Chem. B
{\bf 102}, 10894 (1998).

\bibitem{wan} Q. Wang and J. K. Johnson, J. Chem. Phys. {\bf 110}, 577
(1999).

\bibitem{ham} N. Hamada, S. Sawada, and A. Oshiyama, Phys. Rev. Lett. {\bf
68}, 1579 (1992).

\bibitem{iji2} S. Ijima and T. Ichihashi, Nature {\bf 363}, 603 (1993).

\bibitem{sil1} I. F. Silvera and V. V. Goldman, J. Chem. Phys. {\bf 69},
4209 (1978).

\bibitem{chen} E. Cheng and K. B. Whaley, J. Chem. Phys. {\bf 104}, 3155
(1996).

\bibitem{sil2} I. F. Silvera, Rev. Mod. Phys. {\bf 52}, 393 (1980).

\bibitem{wag} M. Wagner and D. M. Ceperley, J. Low Temp. Phys. {\bf 94},
161 (1994).

\bibitem{sie} Q. Wang, S. R. Challa, D. S. Sholl, and J. K. Johnson,
Phys. Rev. Lett. {\bf 82}, 956 (1999).



\end{thebibliography}
\end{document}